\documentclass[showpacs,superscriptaddress,amsmath,amssymb,nofootinbib,twocolumn,floatfix,reprint]{revtex4-1}

\makeatletter
\def\pdfstartlink@attr{}
\makeatother

\usepackage{multirow}
\usepackage{xcolor}
\usepackage{ulem}

\usepackage{setspace}
\usepackage[hidelinks]{hyperref}
\usepackage{graphicx}
\usepackage[caption=false]{subfig}
\usepackage{lineno}

\newcommand{\td}{{\rm d}}

\begin{document}


\title{Study of scattered light in the main arms of the Einstein Telescope gravitational wave detector}

\author{M. Andr\'es-Carcasona}
\email{mandres@ifae.es (corresponding author)}
\affiliation{Institut de Física d'Altes Energies (IFAE), The Barcelona Institute of Science and Technology, Campus UAB, E-08193 Bellaterra (Barcelona), Spain}
\author{A. Macquet}
\affiliation{Institut de Física d'Altes Energies (IFAE), The Barcelona Institute of Science and Technology, Campus UAB, E-08193 Bellaterra (Barcelona), Spain}
\author{M. Mart\'inez}
\affiliation{Institut de Física d'Altes Energies (IFAE), The Barcelona Institute of Science and Technology, Campus UAB, E-08193 Bellaterra (Barcelona), Spain}
\affiliation{Catalan Institution for Research and Advanced Studies (ICREA), E-08010 Barcelona, Spain}
\author{Ll. M. Mir}
\affiliation{Institut de Física d'Altes Energies (IFAE), The Barcelona Institute of Science and Technology, Campus UAB, E-08193 Bellaterra (Barcelona), Spain}
\author{H.Yamamoto}
\affiliation{LIGO laboratory, California Institute of Technology (Caltech), Pasadena, CA, US}

\date{\today}


\begin{abstract}
 
We present an estimation of the noise induced by scattered light inside the main arms of the Einstein Telescope (ET) gravitational wave detector. Both ET configurations for high- and low-frequency interferometers are considered, for which we propose baffle layouts. The level of scattered light and the ET laser beam clipping losses are intimately related to the baffle inner aperture. We discuss how this translates into minimum requirements on the vacuum pipe radius, a critical parameter in the ET design. The noise estimations are computed using analytical calculations complemented with numerical tools, and depend on a number of baseline parameters we use as input in the calculations. We conclude that the scattered light noise can be maintained at acceptable levels such that does not compromise the ET performance, provided some requirements are met.

\end{abstract}


\maketitle

\section{Introduction}
\label{sec:intro}

The discovery of gravitational waves (GW) in 2015~\cite{FirstGWDet} from a black hole binary merger  by LIGO~\cite{AdvLIGO} started a new era in the exploration of the Universe. The addition of the Virgo antenna~\cite{AdvVIRGO} into the network led in 2017 to the detection of a neutron star binary merger that could be followed in electromagnetic signals, representing the beginning of multi-messenger astronomy.
The Advanced LIGO, Advanced Virgo and KAGRA \cite{KAGRA} interferometers have detected in the O1-O3 observation periods  up to a total of 90 events corresponding to mergers of compact objects~\cite{Abbott2019GWTC-1:Runs,Abbott2020GWTC-2:Run,Abbott2021GWTC-3:Run}. This number is expected to increase in the following joint observation runs, when the upgraded LIGO and Virgo detectors operate with improved sensitivity~\cite{AdvLIGOplus,AdvVIRGOplus}.  Further improvements are being planned to reach the ultimate sensitivity of the current infrastructures in the next few years. The GW community on ground based interferometry is planning for third generation experiments to bring the GW field to the era of precision physics increasing the sensitivity and the rate of detections by orders of magnitude.  Nowadays, the two larger projects under consideration are the Cosmic Explorer in USA~\cite{CE1,CE2,CE3} and the Einstein Telescope (ET)~\cite{ETcds,ETdesign} in Europe.  The results on this paper are based on ET. 

The ET experiment, in its current design, is  envisaged as an underground triangular-shaped experiment with 10~km of arm length, formed by two sets of three nested interferometers optimized at different frequency ranges. In total,  ET will be constituted by six independent Fabry-P\'erot (FP) resonators.   
The low-frequency (ET-LF) triangular configuration uses a 1550~nm laser,  includes cryogenics for cooled mirrors operating at 10 Kelvin, and stores  about 18~kW of laser power in the FP optical cavities.  The high-frequency (ET-HF) triangular configuration uses a laser of 1064~nm, operates at room temperature,  and accumulates about 3~MW of power in the optical cavities.  Therefore, the optical characteristics are rather different and the studies in this paper are performed separately for each configuration.   

The proper definition of the vacuum pipe hosting the laser beam in the main arms constitutes a major milestone in the design of the whole experiment, since it introduces strong constrains over the rest of the detector elements and, to a large extent, conditions the total cost of the project. Therefore,  it is of uttermost importance to perform a thorough study to ensure that the proposed solution fulfills all the requirements such that it does not compromise the projected sensitivity of the experiment (for details on the anticipated ET sensitivity see Refs.\cite{Hild11,NewETsensitivity}).

One of the most important aspects to take into account in establishing the parameters of the vacuum pipe is the control of stray light propagating in the optical cavities, which, if not properly addressed,  has the potential to limit the interferometer sensitivity as a whole. To this end, the vacuum tube is equipped with baffles to absorb the stray light to  avoid 
the couplings with the interferometer's main mode. In this paper, we study the various sources 
of noise due to scattered light inside the main arms of ET and list a number of actions to mitigate or suppress their impact. This naturally leads to a discussion on the
required apertures and the beam pipe radius, as it is closely related to the induced scattered light noise levels.

The work is based on analytical calculations and the results of the pioneering work carried out at LIGO (\cite{Thorne89,Flanagan94}) and Virgo (\cite{Brisson98}).
In addition, numerical simulations of the light fields inside the interferometer are performed using  the \textit{Static Interferometer Simulations} (SIS) software package \cite{Romero21}, which is 
a Fast Fourier Transform (FFT) MATLAB code. As discussed below, this allows for a more refined estimation of the scattered light noise, specially in the case of the backscattering of light out of the baffles inside the vacuum pipe. We use default parameters for the ET-LF and ET-HF optical configurations, as detailed in Refs.\cite{ETcds,ETdesign} and collected in Table~\ref{tab:GeneralParams}\footnote{Here we consider a cartesian right-handed coordinate system with origin in the input mirror of the optical cavity and the $z$-axis along the laser beam line. The azimuthal angle $\varphi$ is measured around the beam axis and the polar angle $\theta$ is measured with respect to the $z$-axis.}.
 
\renewcommand\arraystretch{1.1}
\begin{table}[htb]
\begin{center}
\footnotesize
\begin{tabular}{c |c|c |c|l}
\hline \hline
\multicolumn{5}{c}{FP optical cavity parameters} \\ \hline 
{Variable} & {ET-HF} &{ET-LF} & {Units} & {Description} \\ \hline 
$m$ & $200$ & $211$ & [kg] & Mirror mass \\ \hline 
$L$ & $10$ & $10$ & [km] & Length of an arm \\ \hline 
$\lambda$ & $1064$ & $1550$ & [nm] & Wavelength of the laser \\ \hline
$R_m$ & $0.31$ & $0.225$ & [m] & Radii of the mirrors \\ \hline
$\mathcal{R}_1$  & $5070$ & $5580$ & [m] & Radius of curvature input mirror\\ \hline
$\mathcal{R}_2$  & $5070$ & $5580$ & [m] & Radius of curvature end mirror\\ \hline
$P_{circ}$ & $3000$ & $18$ & [kW] & Circulating power in the cavity \\ \hline
$R$  & $0.5$ & $0.5$ &[m] & Radius of the vacuum pipe. \\ \hline \hline
\end{tabular}
\end{center}
\caption{Parameters of the ET main FP arm cavities used throughout this work. 
The values are extracted from Refs.~\cite{ETcds,ETdesign}.}
\label{tab:GeneralParams}
\end{table}

%
%
\section{Notes on the tube optical aperture and total radius}
 \label{sec:tube}

One of the aspects that conditions the 
size of the vacuum tube is the optical aperture required to maintain at negligible levels  the clipping losses inside the cavity.  
In the ET optical cavities, the laser beam has a  Gaussian-shaped  transverse profile, with maximum beam sizes close to the mirrors at both ends of the FP cavity. 
In a configuration with the laser beam centered inside the vacuum tube, the power decreases exponentially with the distance to the tube longitudinal axis, and the 
beam losses are dictated by the finite apertures inside the vacuum tube.  

\begin{figure}[htb]
\centering
\subfloat[][ET-HF]{\includegraphics[width=\columnwidth]{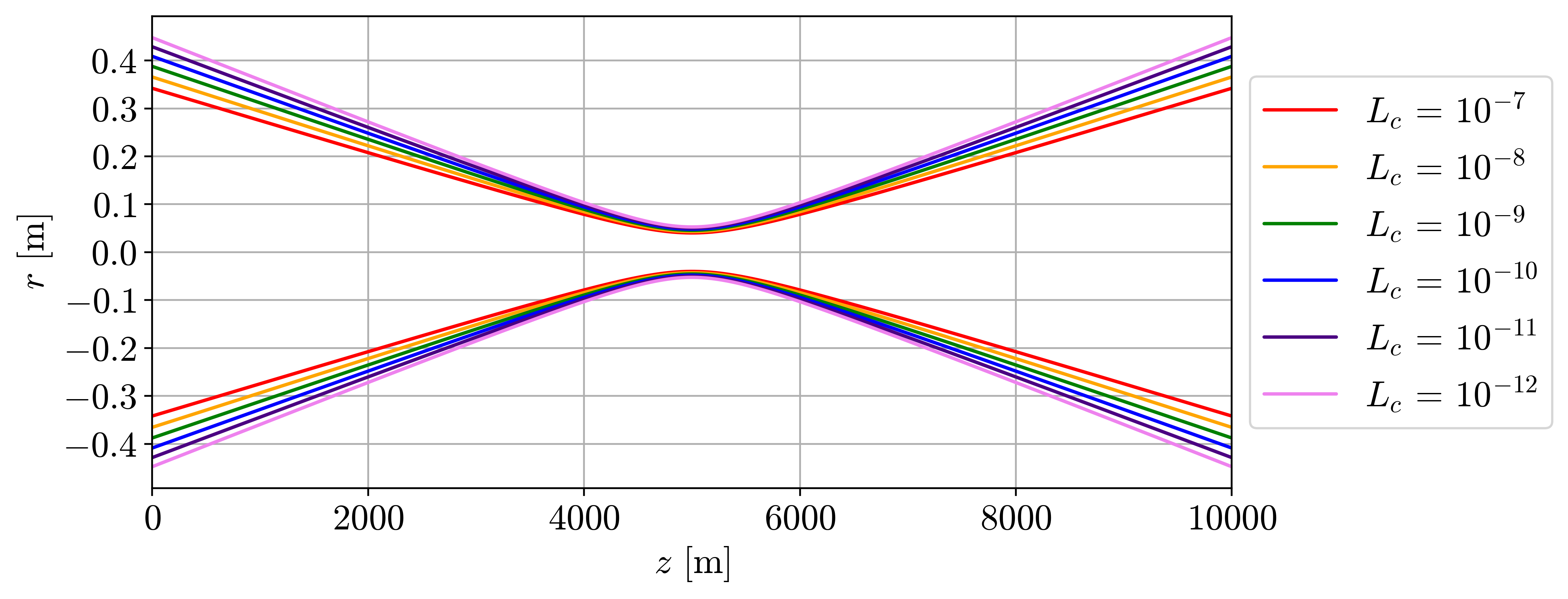}}\\
\subfloat[][ET-LF]{\includegraphics[width=\columnwidth]{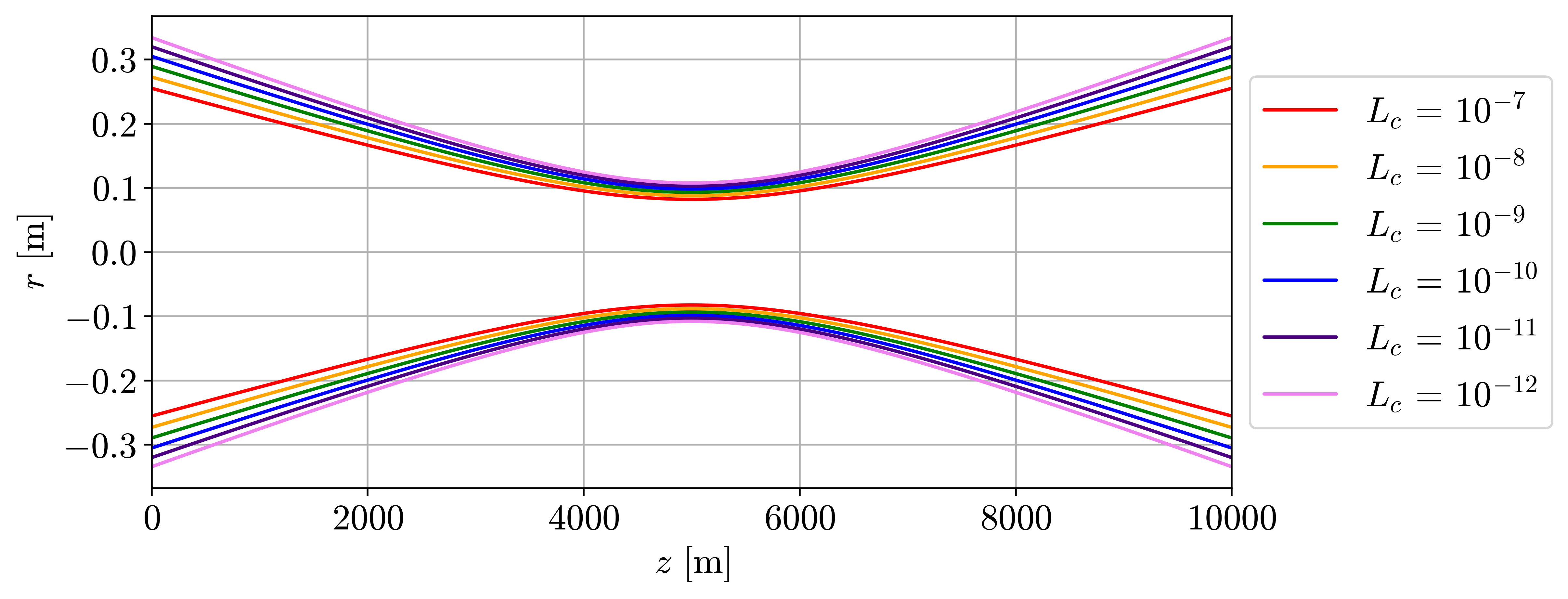}}
\caption{Beam profile as a function of the position in the arm $z$ at different levels of clipping losses $L_c$ for (a) ET-HF and (b) ET-LF interferometers.}
\label{fig:BeamProfileLc}
\end{figure}

The Gaussian beam profile only depends on the optical parameters. The irradiance in cylindrical coordinates is determined by the expression~\cite{Kogelnik66}

\begin{equation}
    I(r,z)= I_0\left( \frac{w_0}{w(z)}\right)^2 \exp\left( \frac{-2r^2}{w(z)^2} \right), 
\end{equation}
where $I_0 = 2P_{circ}/\pi w_0^2$, $P_{circ}$ is the circulating power inside the cavity, $w_0$ the beam waist and $w(z)$ the beam size at position $z$. The beam waist and beam size can be computed from the optical parameters as~\cite{Kogelnik66}
\begin{equation}
    w_0 = \left[ \left(\frac{\lambda}{\pi}\right)^2\frac{L(\mathcal{R}_1-L)(\mathcal{R}_2-L)(\mathcal{R}_1+\mathcal{R}_2-L)}{(\mathcal{R}_1+\mathcal{R}_2-2L)^2} \right]^{1/4}
\end{equation}
and (with $ \mathcal{R}_1 = \mathcal{R}_2$)  
\begin{equation} \label{eq:w(z)}
    w(z) = w_0\sqrt{1+\left( \frac{z-L/2}{z_R} \right)^2}~,
\end{equation}
respectively, with $z_R = \pi w_0^2/ \lambda$, where we adopt the nomenclature for which $z=0$ 
corresponds to the  position of the input mirror in the FP cavity. This expression allows to find the power of the beam inside a radius $r$ at position $z$ as
\begin{equation}
    P(r,z) = \int_0^{2\pi}\int_0^r I(r',z)r'\td r' \td \varphi~, 
\end{equation}
leading to the profile
\begin{equation} \label{eq:Prz}
    P(r,z)=P_{circ} \left[ 1- \exp\left( \frac{-2r^2}{w(z)^2}\right)\right]~.
\end{equation}
\noindent 
Defining the clipping losses as $L_c = 1-P/P_{circ}$, the profile of the beam for a given level of losses 
can be expressed using Eq.~\ref{eq:Prz} as
\begin{equation}\label{eq:r_Lc}
    r(z,L_c) = \frac{w(z)}{\sqrt{2}}\sqrt{\ln \left( \frac{1}{L_c}\right)} + r_{\rm offset}, 
\end{equation}
\noindent
where $r_{\rm offset}$ is an additional term to include eventual offsets of the beam with respect to the cavity 
longitudinal axis. 
Following the expressions above, Fig.~\ref{fig:BeamProfileLc} presents iso-losses curves as a function of the $z$-position in the optical cavity for both ET-HF and ET-LF interferometers, with $r_{\rm offset} = 0$. As expected, the largest beam radius is reached near the mirrors. In a vacuum tube design with constant radius, the required minimal apertures are determined by the losses close 
to the mirrors.  We adopt the criteria that the clipping losses will be maintained at the level $L_c < 10^{-8}$. Evaluating $r(z=0,L_c=10^{-8})$ yields $r_{\mathrm{ET-HF}} \simeq 0.42~\mathrm{m}$ and $r_{\mathrm{ET-LF}} \simeq 0.31~\mathrm{m}$, after  including a beam offset of $r_{\rm offset} = 0.05$~m. These numbers represent the minimum possible inner apertures for the ET-HF and ET-LF interferometers. As discussed in 
Sec.~\ref{sec:baffle}, baffles are installed inside the vacuum pipe to prevent noise due to scattered light from the mirrors reaching the inner walls of the vacuum pipe and re-coupling to the interferometer main laser mode. Therefore, the figures above also represent the minimum baffle inner apertures. 

The requirement of maintaining the clipping losses in the range $L_c < 10^{-8}$ could be regarded as too conservative. 
However, if the baffle inner aperture is chosen to be too narrow, harmful effects beyond the scattered light might appear. For example, diffraction losses caused by the finite inner aperture could increase, reducing the power and efficiency of the entire cavity;  a mode mismatching could appear, as the transverse mode might not properly overlap with the cavity resonating mode,  increasing the losses and the noise of the interferometer; a large sensitivity of the cavity to large misalignment and baffle transverse motions due to microseisms could be introduced; thermal effects due to a larger than expected exposure of the baffle surfaces to the 
light in the cavity might appear together with instabilities in the cavity caused by the change of the optical paths of the photons that hit the baffles; and the sensitivity to any optical imperfection in the mirrors (roughness, and possible defects and contamination) would be increased. Some of these effects are difficult to compute and constitute inherent potential risks for the ultimate sensitivity  of the experiment.

For a typical baffle vertical height in the range of 0.08~m to 0.1~m \cite{Brisson98,Flanagan94} the required aperture translates into a minimum beam pipe radius of $R_{\mathrm{ET-HF}} \simeq 0.5~\mathrm{m}$ and $R_{\mathrm{ET-LF}} \simeq 0.4~\mathrm{m}$.  
The current ET conceptual design includes a vacuum pipe in the main arms with an inner radius of 0.5~m for both ET-LF  and ET-HF configurations.  As presented above, this meets the requirements (but with no additional margin) in the case of ET-HF, and leaves a large margin (about 0.1~m) in the case of ET-LF. For completeness, additional results for ET-HF,  corresponding to a 0.6~m beam pipe inner radius,  are also presented in the following sections and in appendix~\ref{sec:App12}. 

\section{Baffle configuration}
\label{sec:baffle}

\begin{figure*}[htb]
    \centering
\includegraphics[width=0.8\textwidth]{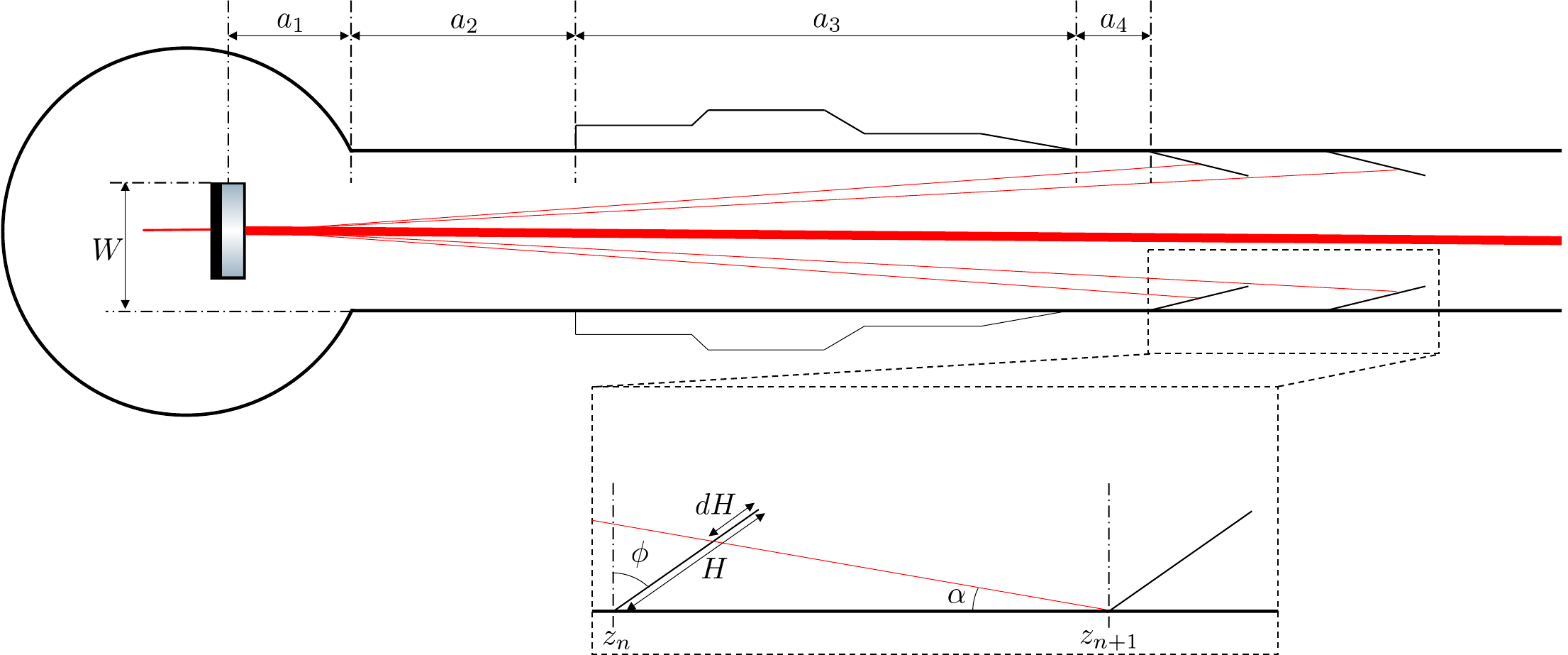}
    \caption{Simplified diagram showing the area of the ET main arm close to the mirror and the geometry and parameters determining the position and shape of the conical baffles (see body of the text). }
    \label{fig:BaffDiag}
\end{figure*}

The currently operating LIGO and Virgo ground-based experiments  implemented a complete set of conical baffles inside the main interferometer arms, with the purpose of mitigating the scattered light noise by geometrically shielding the vacuum tube (and any internal structure on its surface) from photons scattered by the main mirrors in the FP cavities. A simplified scheme of the geometry is  displayed in Fig. \ref{fig:BaffDiag}. Table~\ref{tab:baffles} collects relevant baffle parameters~\footnote{In the case of ET-HF, a vacuum pipe radius of $R = 0.6$~m will be also considered,  leading to $A_b = 1.04$~m}.

\begin{table}[htb]
\begin{center}
\footnotesize
\begin{tabular}{c |c|c |c|l}
\hline \hline
\multicolumn{5}{c}{Baffle parameters} \\ \hline
{Variable} & {ET-HF} &{ET-LF} & {Units} & {Description} \\ \hline 
$A_b$ & $0.84$ & $0.84$ & [m] & Baffle inner aperture  ($R = 0.5$~m) \\ \hline
$A_b$ & $1.04$ & - & [m] & Baffle inner aperture  ($R = 0.6$~m) \\ \hline
$H$ & $0.14$ & $0.14$ & [m] & Baffle length \\ \hline 
$dH$ & $0.0244$ & $0.0244$ & [m] & Baffle overlapping factor \\ \hline 
$\phi$ & $55$ & $55$ & [deg] & Inclination angle of the baffles \\ \hline 
$\displaystyle \frac{\td P}{\td \Omega_{bs}}$ & $10^{-4}$ & $10^{-4}$& [str$^{-1}$] & BRDF  of the baffles \\ \hline \hline
\end{tabular}
\end{center}
\caption{Parameters of the baffles inside the ET vacuum tube used throughout this work (see Fig.~\ref{fig:BaffDiag}).}
\label{tab:baffles}
\end{table}

The distances $a_i$ in Fig.~\ref{fig:BaffDiag} are taken from ET design reports in Refs.~\cite{ET_Tower,ETdesign} and correspond to:

\begin{itemize}
    \item $a_1$: Distance between the mirror and the beginning of the main arm tube [$2.35$ m]. 
    \item $a_2$: Distance between the end of the tower and the beginning of the the cryotrap area [$20$ m]. 
    \item $a_3$: Length of the cryotrap area [$10$ m for ET-HF and $50$ m for ET-LF]. 
    \item $a_4$: Distance between the end of the cryotrap area and the first baffle in the main arm. In this work it will be set to $0$ m, meaning that the first baffle is assumed to be placed just after the cryotrap. 
\end{itemize}
\noindent
With the values for  $a_i$ above,  we establish the position of the first baffle.  These positions are $z_{0} = 32.35$~m and $z_{0}=72.35$~m for ET-HF and ET-LF, respectively. It is important to note this 
work is limited to describe the baffle layout in the bulk of the interferometer's main arms.  No attempt is made to
discuss the stray light mitigation strategy close to the mirrors and inside the cryotrap areas, which constitutes the subject of future studies. 
From Fig. \ref{fig:BaffDiag}, the geometric relation

\begin{equation}
\tan(\alpha) = \frac{W}{z_{n+1}}=\frac{\cos(\phi) (H-dH)}{\sin(\phi)(H-dH)+z_{n+1}-z_n}~,
\end{equation}
\noindent
is extracted, 
which leads to the recursive formula for the baffle position along the $z$-axis 

\begin{equation} \label{eq:BafflePosition}
    z_{n+1}=\frac{W\left[ z_n+ \sin(\phi)(H-dH)\right]}{W-\cos(\phi)(H-dH)}~.
\end{equation}
Setting $\phi=0$, corresponding to baffles perpendicular to the beam line,  one recovers Eq. (3) of Ref. \cite{Flanagan95}.
The distance $W$ represents the farthest point from which a photon should be shielded. We follow 
the conservative approach in Ref.~\cite{Flanagan95} and adopt the maximum possible value corresponding to 
the sum of the beam pipe radius $(R)$ plus half of the radius of the mirror ($R_m$), $W=R+R_m/2$. The expressions above correspond to a baffle layout configuration with the minimum possible number of baffles, and it is driven by pure geometrical arguments. The 
distance between consecutive baffles rapidly increases with  the distance from the mirror such that baffles would be 
mostly placed in the area closer to the mirrors. In practice,  the approach followed by LIGO and Virgo has been that of placing baffles between some of the vacuum pipe sections. This is a more conservative approach from the point of view of scattered light noise mitigation, and it is also motivated by other considerations (going beyond the scope of this publication) related to an efficient massive production and integration of baffles in the vacuum sections,  and the presence of equipment, such as sensors or vacuum pumps, that need to be placed behind baffles to avoid new sources of light scattered noise.

As seen in Fig.~\ref{fig:BaffDiag},  the angle of inclination of the baffle, $\phi$, is chosen such that any photon being scattered at large angles $\theta$ (i.e. from the closest mirror) are deflected at large angles, forcing them to bounce several times before they can reach any of the mirrors. This increases the probability of being absorbed before reaching any of the ends of the vacuum tube. This criterion sets the angle of inclination to be restricted in the range  $\phi\in (45^\circ,90^\circ)$. Lower angles would imply a possible direct back reflection,  severely compromising the sensitivity. In this work we assumed that, with this $\phi$ parameter choice, only the backscattering noise contributions become relevant. The baffle orientation is  symmetric with respect to the middle of the tube. The baffles in the first half of the tube will have $\phi = +55^\circ$ while the ones in the other half of the tube will be inclined with $\phi=-55^\circ$.  The baffles that will receive more light are those exposed to lower $\theta$ values, situated at the far ends of the tube. 

\begin{figure}[htb]
    \centering
    \includegraphics[width=\columnwidth]{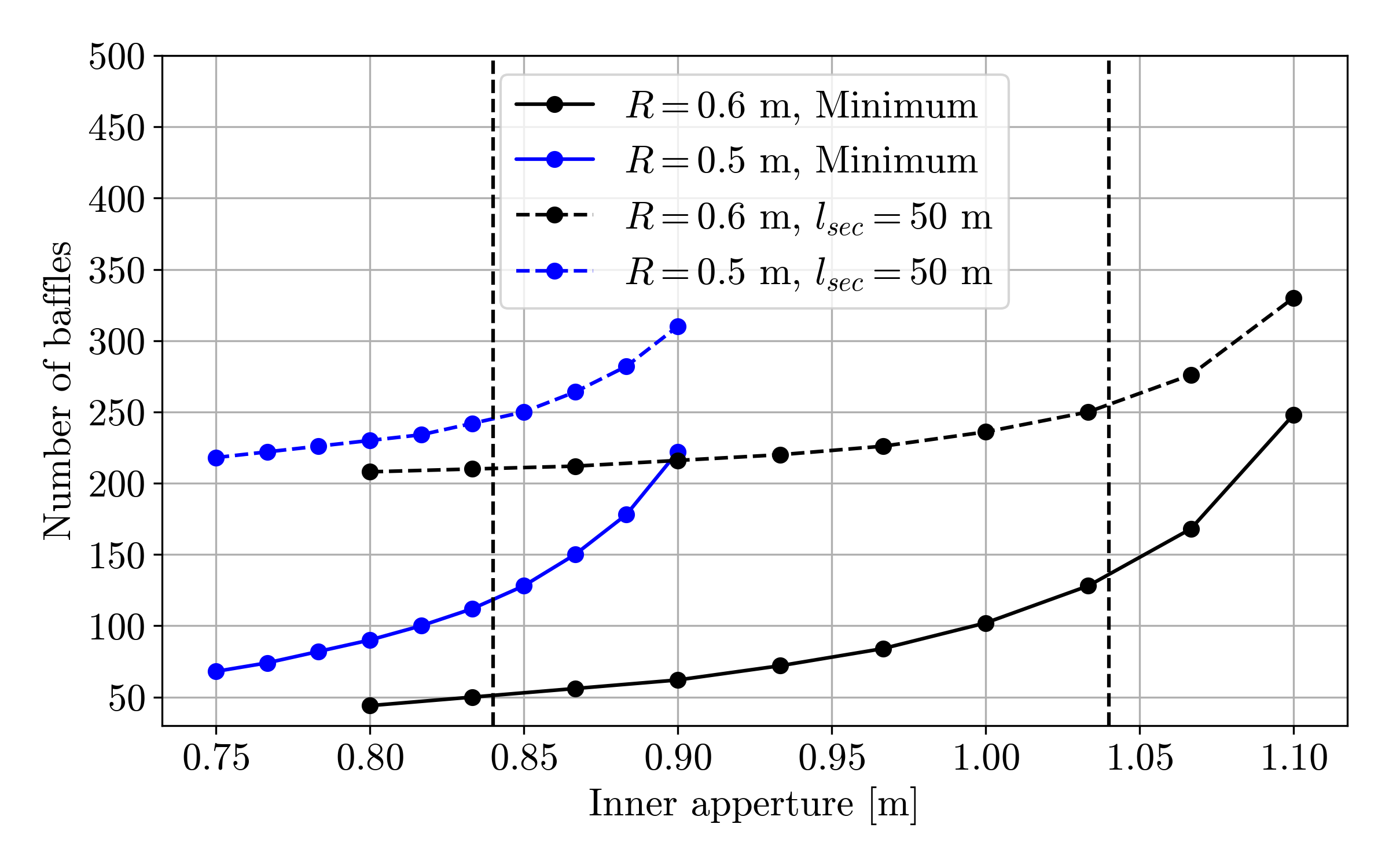}
    \caption{Number of baffles per arm in ET-HF as a function of the baffle inner aperture for a vacuum tube radius of 0.5~m or 0.6~m. The vertical lines indicate the corresponding  inner apertures. 
    Results are provided for both the minimal configuration and after taking into consideration baffles at the end of vacuum pipe sections with a baffle-to-baffle separation of 50~m (see body of the text).}
    \label{fig:NumOfBafflesBafflesHF}
\end{figure}

\begin{figure}[htb]
    \centering
    \includegraphics[width=\columnwidth]{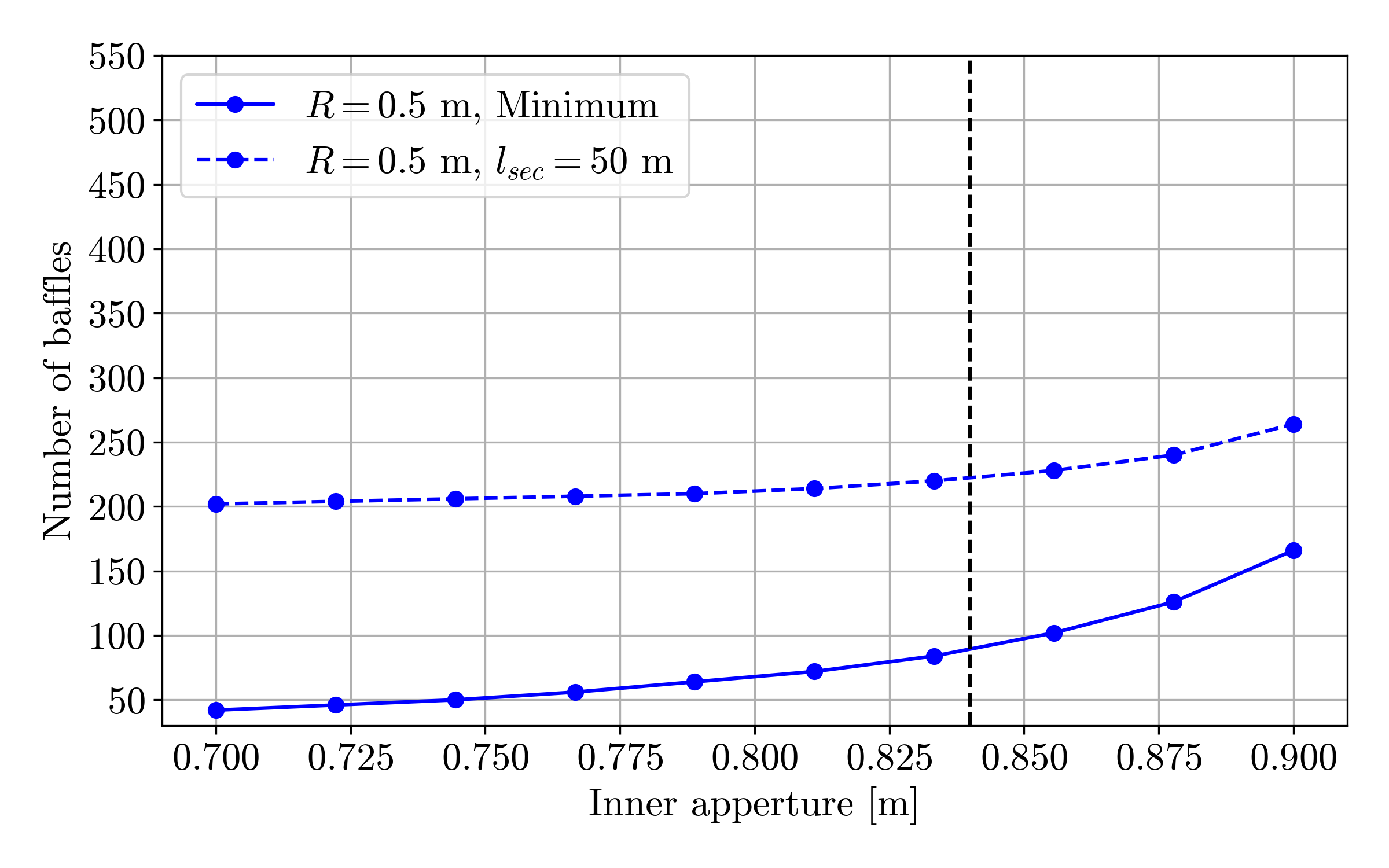}
    \caption{
    Number of baffles per arm in ET-LF as a function of the baffle inner aperture for a vacuum tube radius of 0.5~m. The vertical line indicates the corresponding  inner aperture. 
    Results are provided for both the minimal configuration and 
    after taking into consideration baffles at the end of vacuum pipe sections with a baffle-to-baffle separation of 50~m (see body of the text).}
    \label{fig:NumOfBafflesBafflesLF}
\end{figure}

As discussed in Sec.~\ref{sec:tube}, the inner aperture of the tube and the baffles,  determined as  

\begin{equation}
    A_{b} = 2(R-H\cos(\phi))~,
\end{equation}
\noindent
is a critical parameter that dictates the level of clipping losses in the cavity.  
We compute the required number of baffles as a function of $A_{b}$ separately for ET-HF and ET-LH and in different 
configurations, using the parameters in Table~\ref{tab:baffles}. 
Figure~\ref{fig:NumOfBafflesBafflesHF} presents the results for ET-HF,  either assuming that the minimal 
set of baffles are installed as needed or that baffles are installed as needed close to the mirrors 
and also at the end of some of the vacuum pipe sections.  For the latter we tentatively assume a baffle-to-baffle asymptotic separation, $l_{sec}$, of $50$~m. 
Similarly, Fig.~\ref{fig:NumOfBafflesBafflesLF} shows the results for ET-LH. 
The number of baffles per arm required  in each scenario are collected  in Tab.~\ref{tab:Nbaffles}, where,  in addition and for illustration purposes, results are also presented for a baffle-to-baffle separation of 100~m.  The final baffle-to-baffle separation in ET will depend on the final length of the individual vacuum pipe sections and the very details in the distribution of the vacuum pipe services along the tube.  

\begin{table}[htb]
    \begin{center}
    \footnotesize
    \begin{tabular}{c|c|c|c}
        {ITF} & {Minimum} & $l_{sec}=50 $~m & $l_{sec}=100 $~m  \\ \hline \hline
        {ET-HF} ($0.5$m) & $118$ & $244$ & $162$\\ \hline
        {ET-HF} ($0.6$m) &  $134$ & $254$ & $172$\\ \hline
        {ET-LF} & $90$  & $222$  & $136$\\ \hline
    \end{tabular}
    \end{center}
    \caption{Number of baffles for the ET-HF (using $R=0.5$~m and $R=0.6$~m) and ET-LF in different scenarios and assuming different baffle-to-baffle separations (see body of the text).}
    \label{tab:Nbaffles}
\end{table}

In the following sections,  we describe the elements entering the estimation of the 
scattered light noise produced in the arms. 
In this study, we adopt as benchmark the scenario with $244$ baffles per arm for ET-HF and $222$ baffles per arm for the ET-LF (see Table~\ref{tab:Nbaffles}). 


\section{Mirror's quality}
\label{sec:mirror_quality}
The main source of scattered light noise that we consider here is the one generated by the light scattered by the mirrors. This includes two 
contributions: the mirror's finite aperture and its surface aberration. Any deviation from the perfect surface  causes photons to follow a different path than that intended. To study this effect and characterize the mirror surface defects, mirror maps or phasemaps are used. These are  $n\times n$ arrays containing the height deviation from the perfect spherical surface. They are defined by a resolution $\Delta  =  R_m/n$, which will limit the maximum scattering angles that can be resolved.
The mirror phasemap is measured by a Zygo Fizeau interferometer, with a nominal resolution in the range $0.2 - 0.4$~mm.

A measure of the surface map quality is the spatial power spectral density (PSD). A one-dimensional PSD is obtained from the two-dimensional Fourier transform as explained in Refs. \cite{bass1995handbook,HiroPSD}. Denoting a spatial frequency of this one-dimensional PSD by $\xi$, the scattering angle associated to it takes the form $\theta \sim \lambda \xi$~\cite{stover2012optical}. 
The maximum spatial frequency that can be resolved from the mirror map is the Nyquist frequency, $\xi_\mathrm{max}=1/(2\Delta)$, while the minimum is $\xi_{\mathrm{min}}=1/R_m$ \cite{KAGRA_filtercavity}. The scattering angles relevant for our study are the ones established by the first and last baffles in the vacuum pipe 
\begin{equation} \label{eq:ThetaMaxThetaMin}
    \theta_{\mathrm{max}} = \frac{R}{z_{1}}\sim 10^{-2}~\mathrm{rad},~~~~\theta_{\mathrm{min}} = \frac{R}{z_{N_B}}\sim10^{-5}~\mathrm{rad}~, 
\end{equation}
\noindent
which correspond to spatial frequencies (for $\lambda = 1064~\mathrm{nm}$) 
    $\xi_{\mathrm{min}} \sim 10^{1}~\mathrm{m}^{-1}$ and $\xi_{\mathrm{max}} \sim 10^{4}~\mathrm{m}^{-1}$.
As an approximation, and since the scattered light is larger for lower angles, we compute the contribution of the far half of the tube and double the result, following the approach adopted in Ref.~\cite{Hiro_ETMripple}. 

\begin{figure*}[htb]
    \centering
    \includegraphics[width=0.8\textwidth]{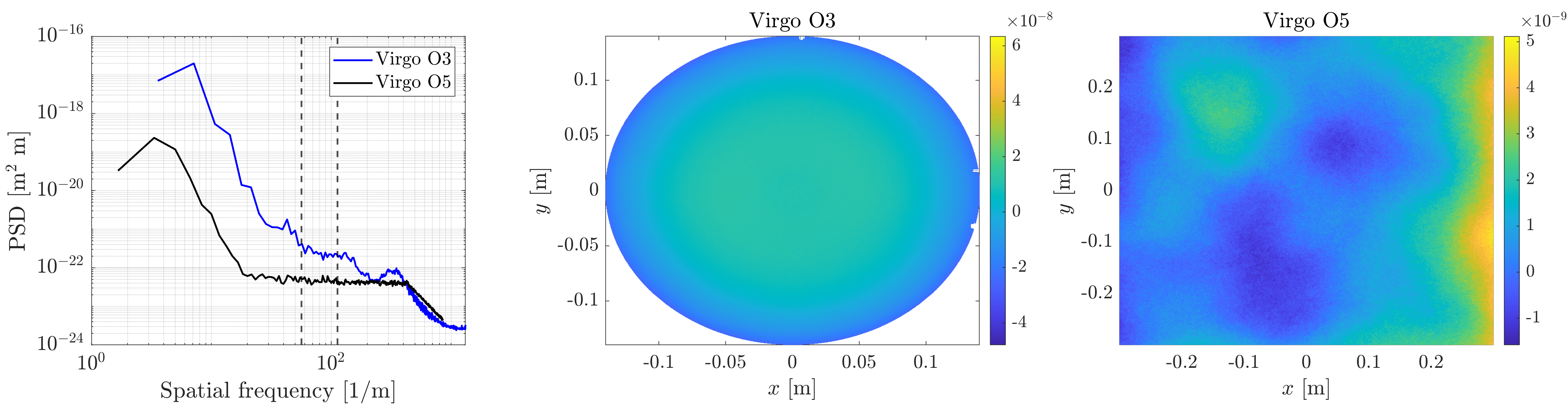}
    \caption{(left)One-dimensional PSDs of Virgo O3 and the projected values for O5 mirror maps. Vertical dashed lines correspond to the spatial frequencies associated to $z=L$ and $z=L/2$ having assumed $\lambda=1064~\mathrm{nm}$. (middle) Height of the deviation from a perfect surface in meters for the north end arm of Virgo during O3. (right) Height of the deviation from a perfect surface in meters projected for O5.}
    \label{fig:PSD}
\end{figure*}

The results in this paper are based on projected mirror maps taken from the Advanced Virgo experiment, corresponding to new mirrors with improved quality, that are  being fabricated and are planned for their installation in Virgo in time for the O5 observation run. The mirror map is constructed from a target PSD set by LMA. 
The same projected mirror maps have been used in recent scattered light noise studies in Virgo~\cite{romero2022determination,Macquet:2022simsVirgo}. Figure~\ref{fig:PSD} presents the projected mirror maps,  and the one-dimensional PSD compared to that of the existing Virgo mirrors used during the O3 observation run.  As clearly observed in the comparison between the measured O3 and the predicted O5 mirror maps, the latter should be regarded as a rather simplified model and constitutes a source of  uncertainty in the scattered light noise calculations.    

We estimate the scattering that each mirror would produce using the bidirectional reflectance distribution function (BRDF), 
defined as the quotient between the differential surface radiance and the surface irradiance,  which takes the form~\cite{stover2012optical}
\begin{equation}
    \mathrm{BRDF} = \frac{1}{P_i\cos \theta_s}\frac{\td P_s}{\td \Omega_s}~,
\end{equation}
\noindent
where the subscripts $i$ and $s$ denote the incident and scattered quantities, respectively. The quantity $\td P_s/\td \Omega_s$ denotes the differential amount of power that goes into a solid angle $\td \Omega_s$. Since the relevant angles of scattering that are going to be considered in this analysis are $\theta\ll 1$ (see Eq. \ref{eq:ThetaMaxThetaMin}), the approximation $\cos \theta_s\approx 1$ holds, making the BRDF directly proportional to the  $\td P_s/\td \Omega_s$ term. The incident power will be constant and equal to the circulating power inside the cavity. The probability of a photon being scattered from a mirror in a solid angle $\td \Omega_{ms}$ is $\td P / \td \Omega_{ms}=P_i^{-1}\td P_s/\td\Omega_s$ implying that 
\begin{equation}
    \mathrm{BRDF} = \frac{\td P}{\td \Omega_{ms}}. 
\end{equation}
\noindent
In the most generic case, the BRDF is a function of both the polar and azimuthal angles, $\theta$ and $\varphi$, respectively. A common simplification is to only consider the azimuthal dependence. The BRDF and the two-dimensional PSD of the surface, $S(f_x,f_y)$ , are related as \cite{stover2012optical}

\begin{equation} \label{eq:BRDFdef}
    \mathrm{BRDF} = \frac{16\pi^2}{\lambda^4}\cos (\theta_i) \cos (\theta_s) Q S(f_x,f_y)~,
\end{equation}
where $Q$ is the geometric mean of the specular reflectances of the surface measured at the incident and scattered azimuthal angles. Equation~(13) describes
the scenario of one single reflection from the mirror surface with a flat incident beam and does not take into account resonator effects and the beam profile inside the cavity.

\begin{figure}[htb]
    \centering
    \includegraphics[width = \columnwidth]{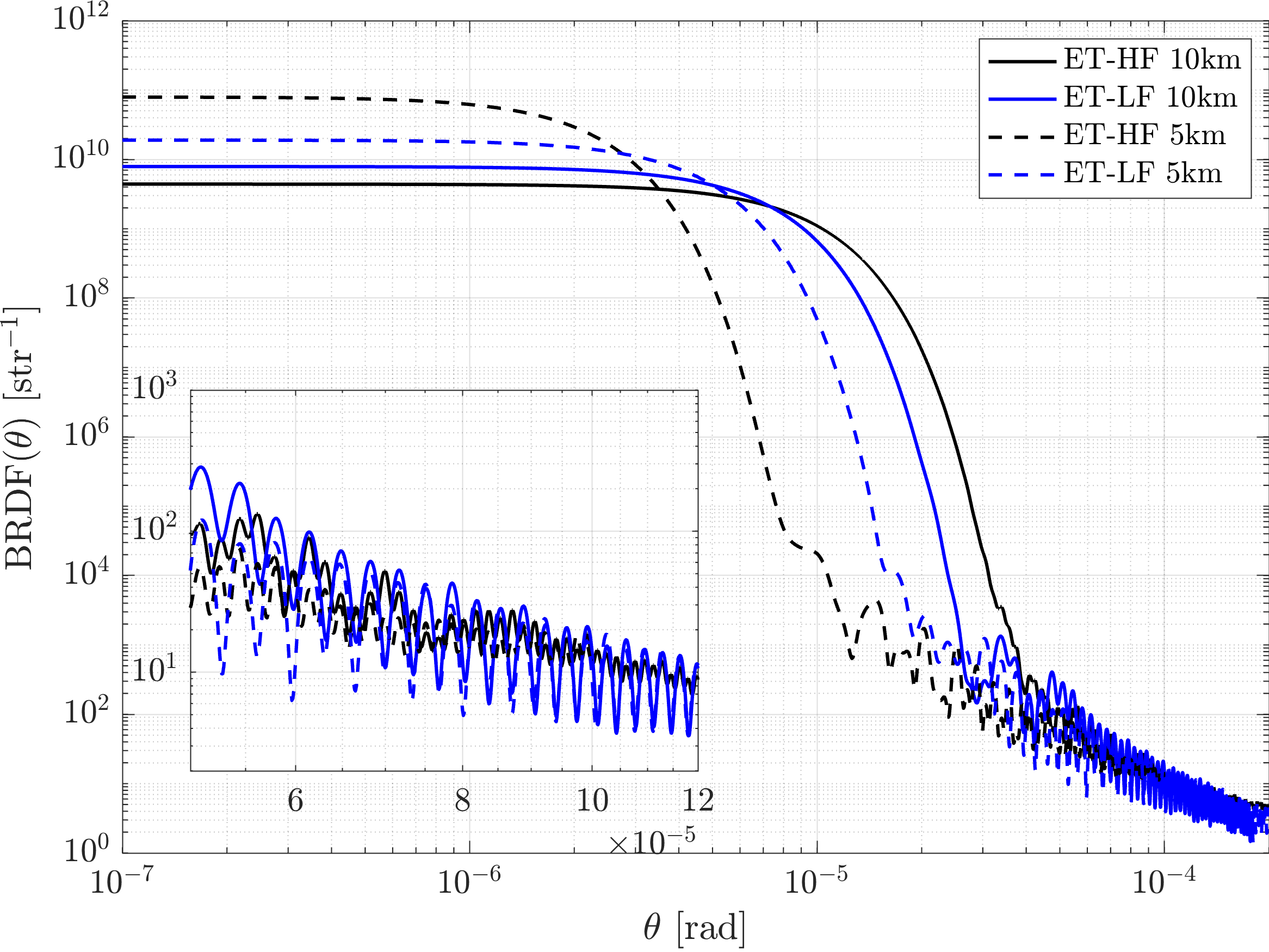}
    \caption{Simulated BRDF, as computed using SIS and using the Virgo mirror maps foreseen for O5 as input, for both the ET-HF and ET-LF. The results are presented at a distance of 10~km (solid lines) and 5~km (dashed lines) from the mirror.
    }
    \label{fig:BRDF}
\end{figure}

The simulated BRDF in the resonating cavity, as determined using SIS, is presented in Fig.~\ref{fig:BRDF} for ET-HF and ET-LF. They are computed at a distance of 5~km and 10~km from a given mirror and are based on the quality of the projected O5 mirror maps~\footnote{The SIS simulation does not include the effect of disturbances in the field inside the cavity due to the presence of the baffles themselves. However, this effect is small and is not expected to change the conclusions.}. The flat BRDF distribution at very low angles is driven by the power distribution in the resonant cavity. In the case of ET-LF, the BRDF distribution shows a clear oscillatory pattern at large angles. This is attributed to aperture effects, with  
the scattering being dominated by the finite size of the mirror (see appendix~\ref{sec:finite}) 
and for which the effect of the mirror's surface aberration is subdominant. Such an oscillatory pattern 
is less pronounced in the case of ET-HF, indicating a major contribution from the mirror's roughness. The BRDF levels obtained at  5~km and 10~km distance from the mirror tend to converge at very large angles,  indicating the power distribution observed by the baffles at those fixed locations is similar.  

\section{Seismic displacement}
\label{sec:seismic}

The photons scattered from the mirrors, away from the main path in the cavity, illuminate the vacuum pipe baffles\footnote{As pointed out, the baffle layout inside the vacuum tube is designed to avoid that photons scattered from any point in the mirror can reach the inner surface of the bare vacuum tube.}. These are vibrating surfaces, due to seismic noise, that introduce changes in the photon's phase, and mimic a gravitational wave signal if the photons  re-couple to the interferometer's main mode. As already pointed out, no attempt is made to include other contributions 
originated from scattering areas in the vicinity of the mirrors or inside the cryotraps, that will require a separate study. 

In general, the relevant quantity is the longitudinal motion of the baffles \cite{Thorne89}. This quantity is related to the seismic motion of the ground, $X(f)$, as 
\begin{equation}
    X_{\mathrm{baff}}(f)=H_{\mathrm{tube}\rightarrow \mathrm{baff}}(f)H_{\mathrm{ground}\rightarrow \mathrm{tube}}(f) X(f)~,
\end{equation}
\noindent
where $H_{\mathrm{ground}\rightarrow \mathrm{tube}}(f)$ denotes the transfer function describing how the ground motion translates into a motion of the vacuum tube, and 
$H_{\mathrm{tube}\rightarrow \mathrm{baff}}(f)$ is the transfer faction describing the 
propagation of the vacuum tube motion to the baffles.  Those transfer functions are not known at the moment but can be simulated and need to be measured once a prototype of the beam pipe is constructed. 
In this work, $H_{\mathrm{tube}\rightarrow \mathrm{baff}}(f)=H_{\mathrm{ground}\rightarrow \mathrm{tube}}(f)=1$ are used.  
A good mechanical design of the coupling between the vacuum tube and the baffles will contribute to  damping  the seismic vibrations and reducing the scattering noise contributions.  

\begin{figure}[htb]
    \centering
    \includegraphics[width = \columnwidth]{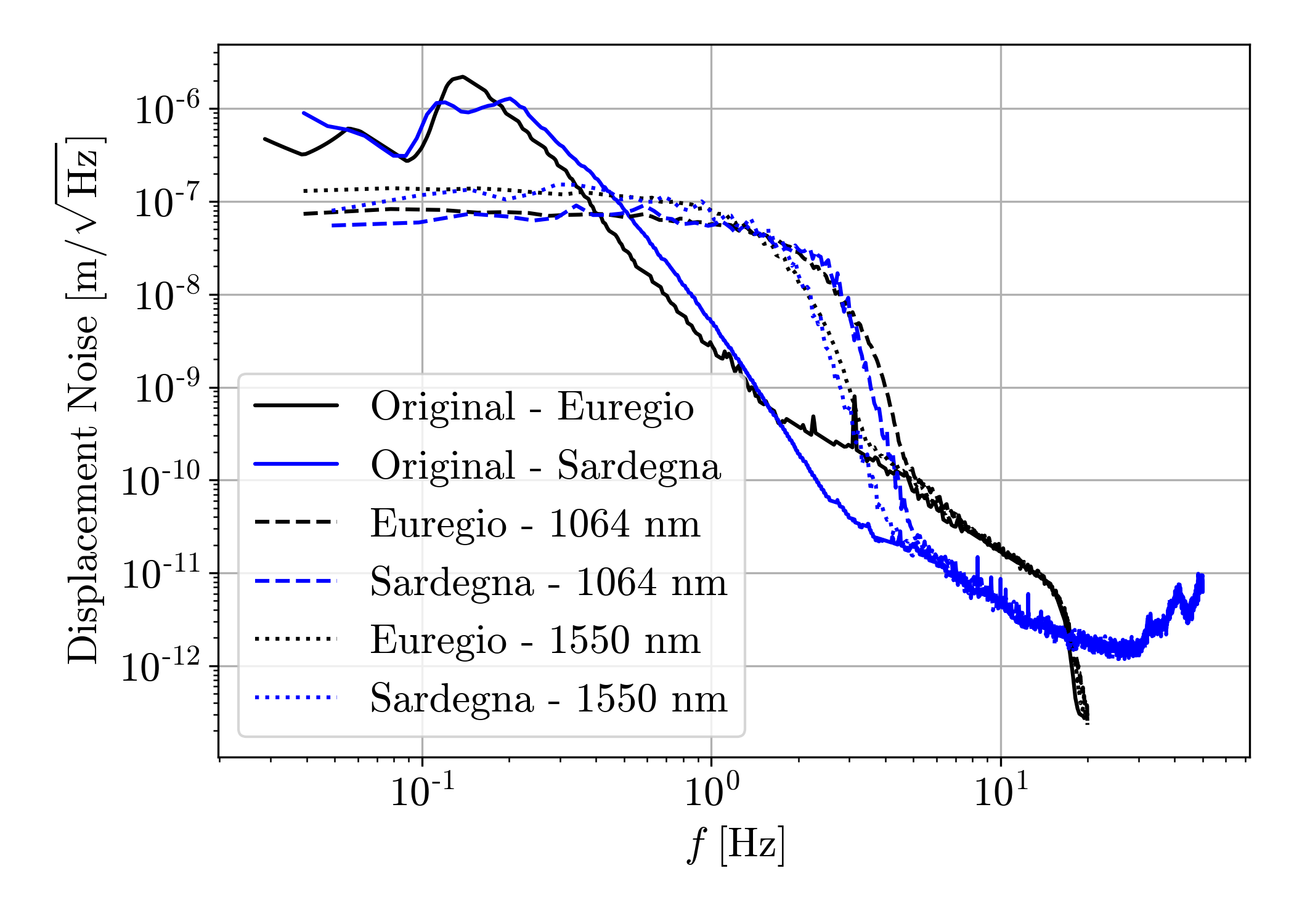}
    \caption{Original seismic noise for the Euregio and Sardegna sites and the upconverted ones for $\lambda = 1064$ nm and $\lambda = 1550$ nm.}
    \label{fig:SeismicNoise}
\end{figure}

Turning now into the seismic noise, in this study we considered two possible locations for hosting ET:  the Sos Enattos mine area in Sardegna~\cite{SoSEnattos1,SoSEnattos2} in Italy, and the Euregio Rhein Maas~\cite{ERM} area in the borders between Belgium, The Netherlands and Germany. Thanks to thorough campaigns carried out  in both sites to characterize the seismic motion, seismic data is available and will be input to our calculations.
The data available take the form of PSDs of either the velocity ($\mathrm{PSD}_{\rm vel}$) or the acceleration noise ($\mathrm{PSD}_{\rm acc}$). The relations in the frequency domain between these noises and the displacement noise ($\mathrm{PSD}_{\rm dis}$) are 
\begin{equation}
    \mathrm{PSD}_{\rm acc}(f) = (2\pi f)^2  \mathrm{PSD}_{\rm vel}(f) = (2\pi f)^{4}  \mathrm{PSD}_{\rm dis}(f)~.
\end{equation}
\noindent
If the seismic displacement is similar to or greater than the optical wavelength, a phenomenon known as phase-wrapping occurs. To account for this effect an upconversion or fringe-wrapping of the noise is performed. As stated in Ref. \cite{CE_backscattering}, this can be computed as
\begin{equation}
    X(f) = \frac{\lambda}{4\pi}\sqrt{\mathrm{PSD}\left[ \mathcal{S}(t) \right]+\mathrm{PSD}\left[ \mathcal{C}(t) \right]}~,
\end{equation}
with
\begin{equation}
    \mathcal{S}(t) = \sin\left( \frac{4\pi}{\lambda} X(t) \right)
\end{equation}
and 
\begin{equation}
    \mathcal{C}(t) = \cos\left( \frac{4\pi}{\lambda} X(t) \right) ~.
\end{equation}
\noindent
The input data must be expressed in time domain. Based on the 
PSD of the seismic noise, we generate a mock time domain data with the same spectrum using
\begin{equation}
    X(t) = \sum_{f=f_{\mathrm{min}}}^{f_{\mathrm{max}}} A(f)\sin(2\pi f t + \Phi)~,
\end{equation}
where $A(f)= \sqrt{2 \mathrm{PSD}(f) \Delta f}$ and $\Phi\sim \mathrm{Uniform}(0,2\pi)$. The frequency vector, $f$, is sampled at a constant step $\Delta f$. The time domain data $t\in[0,(M-1)\Delta t]$ is formed by $M$ samples equally spaced by $\Delta t=1/(M\Delta f)$. Figure~\ref{fig:SeismicNoise} displays the original PSDs at both locations and the resulting PSDs after the noise has been phase-wrapped. As expected, the noise at low frequencies is reduced and the one at large frequencies is enhanced. This result is input as the effective displacement in our scattered light noise calculations. 
As shown in Fig.~\ref{fig:SeismicNoise} the results from Euregio site are limited to frequencies below 20~Hz whereas the Sardegna spectrum extends up to about 50~Hz.    

\section{Analytic study}
\label{sec:analytic}
An estimation of the scattered light noise is performed using an analytical approach. Here, four sources of noise are considered~\cite{Brisson98}:
\begin{itemize}
    \item {\it{Diffraction}}. 
    The limited inner apertures of the baffles produce a 
    diffraction pattern that can limit the sensitivity of the interferometer. This noise contribution depends on the number of baffles placed inside the vacuum tube and their geometry,  as the single noise contributions from each baffle can pile up coherently. The main mitigation strategy is to break this coherence, which can be achieved by serrating the edges of the baffles as has been done in the case of LIGO and Virgo \cite{Brisson98,Thorne89,Flanagan94,Flanagan95,FlanaganThorne95_Diff,Vinet96}. This reduces by about two orders of magnitude its contribution to the scattered light noise budget. Additionally, extra radiation pressure to the test masses can be relevant at low frequencies.
    \item {\it{Backscattering}}. This is caused by photons that after being scattered by one of the mirrors reach another surface, such as a baffle, where they scatter again and reach any of the mirrors. These photons can recombine with the main beam and, bringing a different phase, limit the sensitivity of the interferometer. Again, extra radiation pressure to the test masses become relevant at low frequencies.
    \item {\it{Baffle edges}}. Baffles are inclined exposing to photons their edges. In principle, photons can be reflected on these edges and reach any of the mirrors. A simple solution to virtually eliminate this noise contribution is also that of serrating the baffle edges. 
    \item {\it{Shinning facet}}. In principle, any piece of reflective material inside the vacuum tube can be a source of significant scattered light noise as any strayed photon can be directly reflected back to any of the mirrors. Such  reflective surfaces can be produced during the machining of the tube or by equipment placed inside it. This effect is totally suppressed by placing baffles such that no strayed photon can reach the bare structure of the vacuum tube.
\end{itemize}
\noindent
Following the considerations above, only the backscattered and diffraction noise contributions will be further analyzed. The rest of noise sources are orders or magnitude smaller, provided the recommendations related to the baffle layout inside the vacuum tube and the serration of the baffle edges are followed.

\subsection{Diffraction noise}
The finite inner aperture produces in each baffle a diffraction pattern that adds up coherently when the baffle has smooth edges. As stated in Ref. \cite{FlanaganThorne95_Diff}, the noise generated by the clear aperture of smooth baffles in a centered mirrors cavity can be expressed as  

\begin{equation}
    \tilde{h}_{\mathrm{smooth}}(f)=\frac{\kappa \lambda X(f)}{\sqrt{2}LR}\sqrt{N_{B}}~,
\end{equation}
\noindent
where the contributions from $N_B$ baffles are added with no interference, 
and $\kappa$ is a parameter extracted from a fit to the tail of the mirror's BRDF using
the functional form $\mathrm{BRDF}=\kappa\theta^{-2}$ for $\theta \in [\theta_{\min}, \theta_{\max}]$. The values obtained for $\kappa$ are 
$\kappa=1.32\times 10^{-7}$ str$^{-1}$ for ET-HF and $\kappa=1.58\times 10^{-7}$ str$^{-1}$ for ET-LF. 

As already pointed out, the solution adopted by LIGO and Virgo to mitigate the diffraction  noise is to break the coherence in the diffraction pattern by adding a particular triangular serration to the inner edge of the baffles.
The peak to valley serration height, denoted by $\Delta H$, must be large when compared to the width of a Fresnel zone \cite{FlanaganThorne95_Diff},
\begin{equation}
    \mathcal{W}(z)=\frac{\lambda}{2LR}z(L-z)~.
\end{equation}
\noindent
The maximum Fresnel zone is found at $z=L/2$ and equals $\mathcal{W}_{\max}= \frac{\lambda L}{8R}$ which translates into values of  $2.66$~mm  and $3.88$~mm for ET-HF and ET-LF, respectively. A peak to valley serration of $1$ cm is chosen to ensure that $\Delta H/\mathcal{W}\gg 1$.
In order to  completely break the coherence this peak to valley height should be randomized by an amount $\gtrsim 2\mathcal{W}$. This translates into a final peak to valley height of $\Delta H = \overline{\Delta H}+\mathrm{Uniform}(-\mathcal{W},\mathcal{W})$ where the mean height, $\overline{\Delta H}$, is kept as $1$ cm and $\mathcal{W}=4$ mm is chosen to fulfill the requirements for the most restrictive case, corresponding to the baffle placed at the middle of the vacuum tube in the ET-LF configuration.
Finally, this serration criteria imposes a lower bound on the baffle length safety margin, $dH$, as the valleys of the serration could allow photons to reach the tube if the condition 
\begin{equation}
    dH\geq \frac{\overline{\Delta H}+\mathcal{W}}{\cos(\theta)}
\end{equation}
is not fulfilled. 
Altogether, this results into setting the value $dH = 2.44$ cm in Table~\ref{tab:baffles}.

The level of diffraction noise in the case of serrated baffles, as defined above, takes the form \cite{FlanaganThorne95_Diff}
\begin{equation}
   \tilde{h}_{\mathrm{random}}(f)=\frac{\kappa \lambda X(f)\sqrt{N_B}}{LR}\left[ \frac{\lambda L}{8\pi R\overline{\Delta H}}\right]\left[\frac{\sqrt{\lambda L/4}}{2\pi R}\right]^{1/2}~.
\end{equation}
\noindent
To account for the radiation pressure on the mirrors, the diffraction noise must be modified by
\begin{equation}
       \tilde{h}_{\mathrm{diff}}(f)= \tilde{h}_{\mathrm{smooth/random}}(f) \sqrt{1+\left(\frac{8 \Gamma P_{circ}}{cM\pi f^2}\right)^{2} \frac{1}{\lambda^2}}, 
\end{equation}

\noindent
where $\Gamma$ represents the gain of the cavity formed by the FP input mirror and the signal recycling mirror~\cite{Hiro_ETMripple} and $M$ is the mass of the mirror. 
The value for $\Gamma$ is estimated as in Ref.~\cite{Hiro_ETMripple}
\begin{equation}
    \Gamma = \frac{1-r_{i}r_{s}}{1-r_{i}r_{s}-r_{i}+r_{s}}~,
\end{equation}
\noindent
with $r_i$ and $r_s$ being the amplitude reflectivity of the input mirror and the signal recycling mirror, respectively. 
In this study we adopt the value $\Gamma = 15.7$.
Equation~(24) is  used to determine the diffraction noise levels for ET-HF and ET-LF configurations.

\subsection{Backscattering}

In order to quantify the noise contribution related to this effect we first follow the work in Refs.~\cite{Thorne89,Flanagan94,Flanagan95,FlanaganThorne95_Diff}, with 
the differential strain noise defined as
\begin{equation}
    \td \tilde{h}^2(f)=\left( \frac{\lambda}{z}\right)^{2}\left( \frac{\td P}{\td \Omega_{ms}} \right)^2\frac{\td P}{\td \Omega_{bs}} X^2(f)\delta \Omega_{ms}~,
\end{equation}
\noindent
where $z$ is the distance from one mirror to the position where the backscattering takes place (for example the baffle locations), $\td P/\td \Omega_{bs}$ represents the probability of a photon of being scattered by this surface and $\delta \Omega_{ms}$ is the solid angle seen by the photon being scattered off the mirror (see Eqs. (18) and (28) in Ref. \cite{Flanagan95}). 
The latter expression only considers phase noise contributions and neglects the amplitude noise related to the radiation pressure generated by the backscattered photons. As shown in Ref.~\cite{Hiro_ETMripple}, this effect might not be negligible at low frequencies. In order to 
take into account this contribution, the differential strain noise expression is modified as
\begin{equation} \label{eq:diff_strain}
    \td \tilde{h}^2(f)=\frac{1}{L ^2}\left[\lambda^2+\left(\frac{8 \Gamma P_{circ}}{cM\pi f^2}\right)^{2}\right]\frac{\td P}{\td \Omega_{bs}} X^2(f)\td K~,
\end{equation}
\noindent
with 
\begin{equation}
    \td K = \frac{1}{z^2}\left( \frac{\td P}{\td \Omega_{ms}} \right)^2\delta \Omega_{ms}~.
\end{equation}
\noindent
Integrating Eq. \ref{eq:diff_strain} one obtains
\begin{equation} \label{eq:hBackscatteringFinal}
    \tilde{h}^2_{\mathrm{back}}(f)=\frac{1}{L^2}\left[\lambda^2+\left(\frac{8 \Gamma P_{circ}}{cM\pi f^2}\right)^{2}\right]\frac{\td P}{\td \Omega_{bs}} X^2(f) \int_{\theta_{\mathrm{min}}}^{\theta_{\mathrm{max}}} \td K~.
\end{equation}
\noindent
In  Refs.~\cite{Thorne89,Flanagan94,Flanagan95} the integral is performed assuming that the solid angle is $\delta \Omega_{ms}=2\pi\theta \td \theta$. This approach ignores the shielding among two consecutive baffles. Instead, we perform the integral as
\begin{equation}
    \int_{\theta_{\mathrm{min}}}^{\theta_{\mathrm{max}}} \td K\simeq \sum_{i=1}^{N_B}
    \frac{1}{z_i^2}\left( \frac{\td P^i}{\td \Omega_{ms}}\right)^2\delta \Omega_{ms}^i~. 
\end{equation}
\noindent
The term $z_i$ is the position of each baffle, the sum runs over the number of baffles $N_B$, and the effective solid angle for each baffle is
\begin{equation}
    \delta \Omega_{ms}^i = 
    \begin{cases}
        \displaystyle \frac{2\pi H\cos \phi}{z_i^2}(R-H\cos \phi) & \text{if } i=1\\ \\
        \displaystyle \frac{2\pi (R-H\cos \phi)^2}{z_i}\left[\frac{1}{z_{i-1}}-\frac{1}{z_i} \right] & \text{if } i\neq 1
    \end{cases}~.
\end{equation}
\noindent
The term $\td P^i/\td \Omega_{ms}$ can be computed analytically, using the BRDF shown in Sec. \ref{sec:mirror_quality}, or via numerical simulations as explained below. 

\section{Numerical simulations}
\label{sec:numerical}

The simulations are performed using SIS. 
The mirror maps in Sec. \ref{sec:mirror_quality} are used to specify the roughness of the surfaces, and the rest of the optical cavity is specified according to the parameters collected in Tab. \ref{tab:GeneralParams}.
We adopt the expression, 

\begin{equation}
    \frac{\td P^i}{\td \Omega_{ms}} \approx  \frac{P^i/P_{circ}}{\delta \Omega_{ms}^i}~,
\end{equation}
\noindent
where the term $P^i$ indicates the amount of power hitting the $i^{\rm th}$ baffle. As explained below, in computing this term,  we add the power reaching the baffle only on the area not shielded by a previous baffle. 

As a FFT-based code,
SIS requires a spatial resolution to be specified,  that dictates the accuracy of the resulting simulations. We use a grid with resolution of $2$ mm,  which allows simulating scattering angles up to  $\theta \approx 5\times 10^{-4}$ rad.  The use of a much smaller spatial resolution would render the numerical calculations unfeasible and computationally very heavy. As a consequence, the power 
illuminating each of the baffles in the cavity from the scattering of a given mirror cannot be computed with the required level of accuracy. In particular, this affects baffles in the first half of the tube, closer to the given mirror, which correspond to larger scattering angles. As already pointed out in Sec.~\ref{sec:mirror_quality}, small angle scattering dominates and the simulations are only performed for the second half of the baffles, far from the mirror, and doubled to take into account the presence of both mirrors in the cavity.  This approach is also supported by the symmetry of the beam profile and by the fact that the tilt of the baffles prevents any direct back reflection to the close-by mirrors.

\begin{figure*}[htbp]
    \centering
\includegraphics[width=0.8\textwidth]{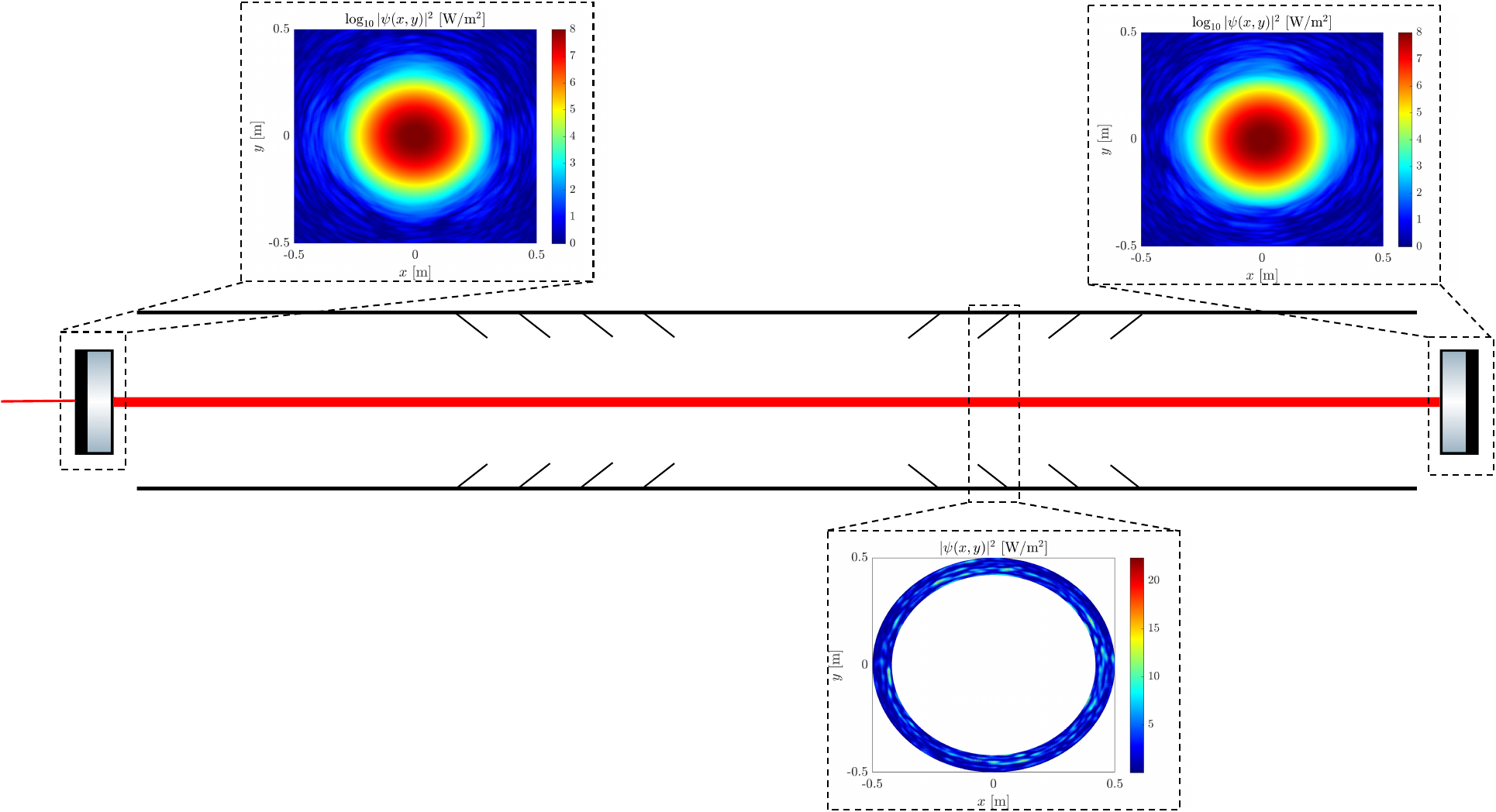}
    \caption{Scheme showing the FP cavity simulated and the power distribution at the FP mirrors and at the position of a baffle at $z=8000$m.}
    \label{fig:SchemeFP}
\end{figure*}

In a locked FP resonating cavity, the circulating power is maximized and the field, $\psi(x,y,z)$, can be computed using SIS at any position.  The power reaching the $i$th baffle follows 
the expression
\begin{equation}
    P^i = \iint_{A^i} |\psi(x,y,z^i)|^2 \td A^i~, 
\end{equation}
\noindent
where the term $\td A^i$ denotes  the differential area $\td A^i = r\td r \td \varphi$. Only the non-shielded region of the baffle is integrated over ($r\in(A_b/2,r_s^i)$ and $\varphi \in (0,2\pi)$),  where 

\begin{equation}
    r_s^i = (R-H\cos(\phi))\frac{z_i}{z_{i-1}}~,  
\end{equation}
\noindent
which takes into account the shielding between two consecutive baffles. 
Numerically, the integrated power is approximated as
\begin{equation}
    P^i\approx \sum_{k}\sum_{j} |\psi_{jk}|^2 A_{jk}~,
\end{equation}
\noindent
where $A_{jk}$ is the baffle area illuminated at each cell of the grid in SIS.  Figure~\ref{fig:SchemeFP}
shows a schematic of the FP cavity,  together with the simulated power distributions 
at the FP mirrors, and at a baffle located at $z=8000$~m.
The Gaussian profile of the light at the core of the mirrors is clearly observed. Random 
structures in the simulated power appear with increasing the transverse 
distance from the center of the laser beam which are dominated by scattered light,  and are  dictated by the details of the mirror maps. In particular, the power that reaches the baffles is totally dominated by the tails in the laser beam generated by scattered light.

\section{Results}
\label{sec:results}

The estimated scattered light noise from diffraction or backscattering contributions are computed following the prescriptions presented in the previous sections.   They are compared to the anticipated ET sensitivity for both ET-HF and ET-LF configurations. 

\begin{figure*}[htbp]
    \centering
    \includegraphics[width=\columnwidth]{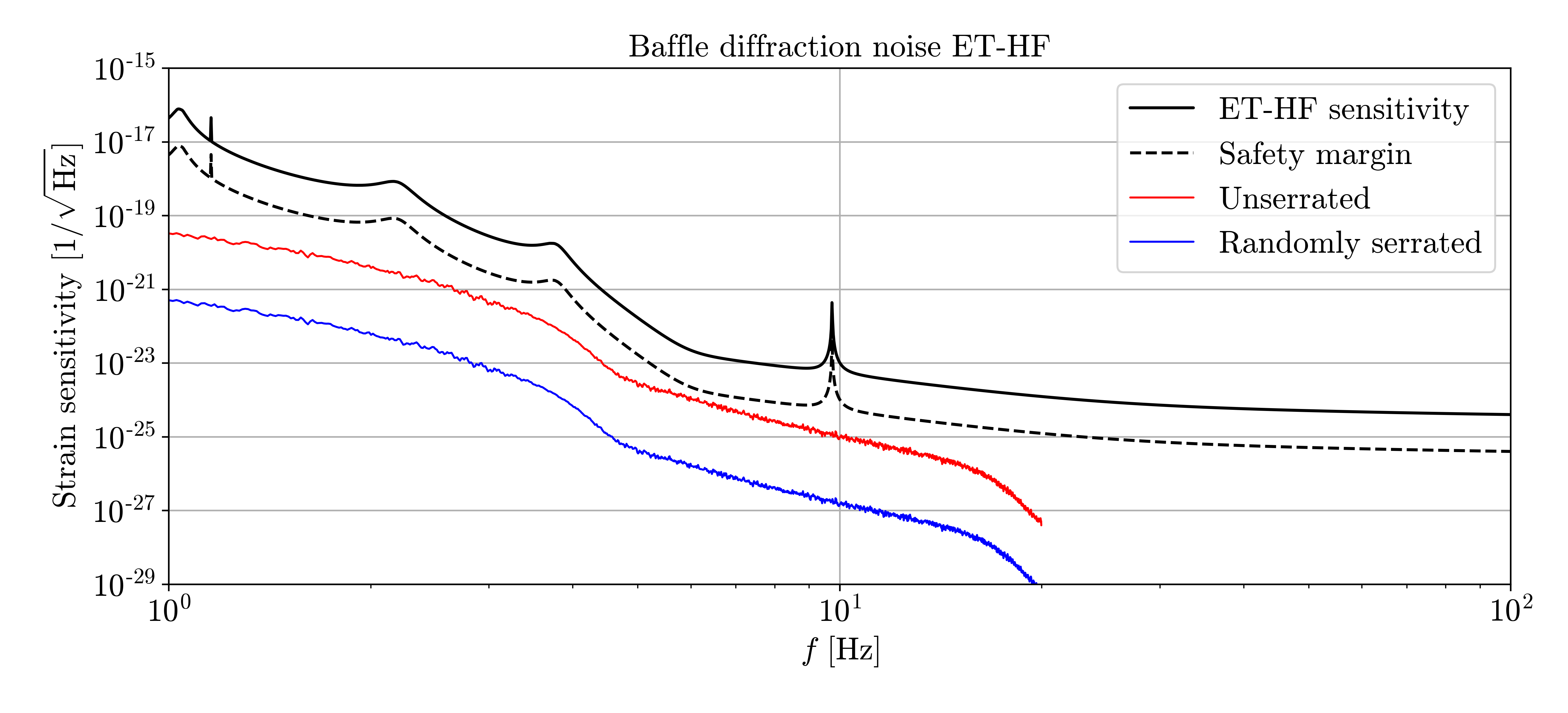}
    \includegraphics[width=\columnwidth]{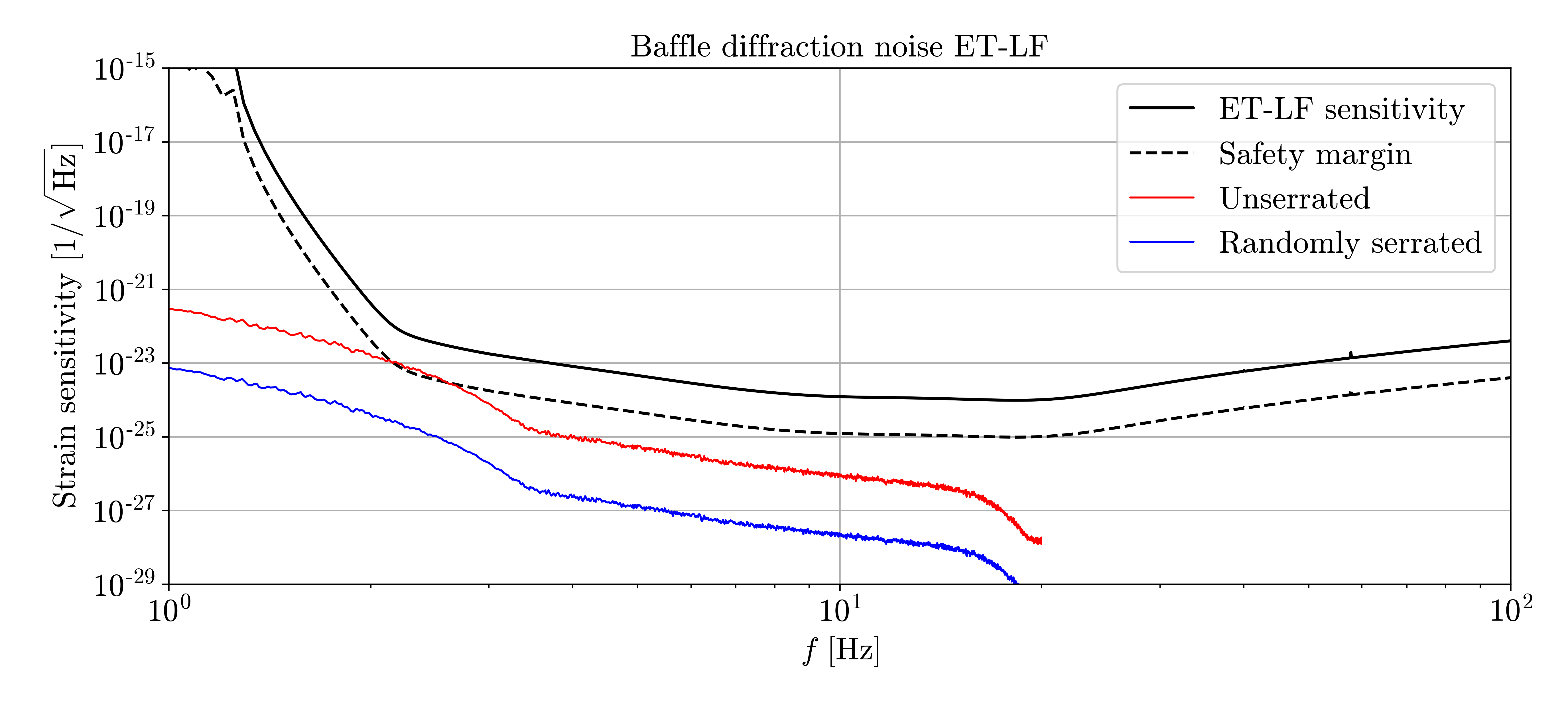}
    \caption{Stray light noise due to diffraction effects as a function of frequency. The results for unserrated (red lines) and randomly serrated (blue lines) baffle edges are compared to the anticipated ET sensitivity (black lines) and the corresponding 1/10 safety margin (dashed lines). The results are presented separately for (left) ET-HF and (right) ET-LF configurations. The seismic noise data from the Euregio site are used.}
    \label{fig:diff_LF}
\end{figure*}

In the case of the diffraction noise contributions, the results are computed adopting Eq.(23) and
the seismic noise levels from the Euregio site. Similar results are obtained if the Sardegna  seismic noise curve is used. For illustration purposes, the results are presented in 
Fig.~\ref{fig:diff_LF} for both smooth (unserrated) and randomly serrated baffle edges, separately for ET-HF and ET-LF.  As already pointed out, the serration of the baffle edges is mandatory and its implementation brings the diffraction noise to acceptable levels.  

\begin{figure}[htbp]
    \centering
    \includegraphics[width=\columnwidth]{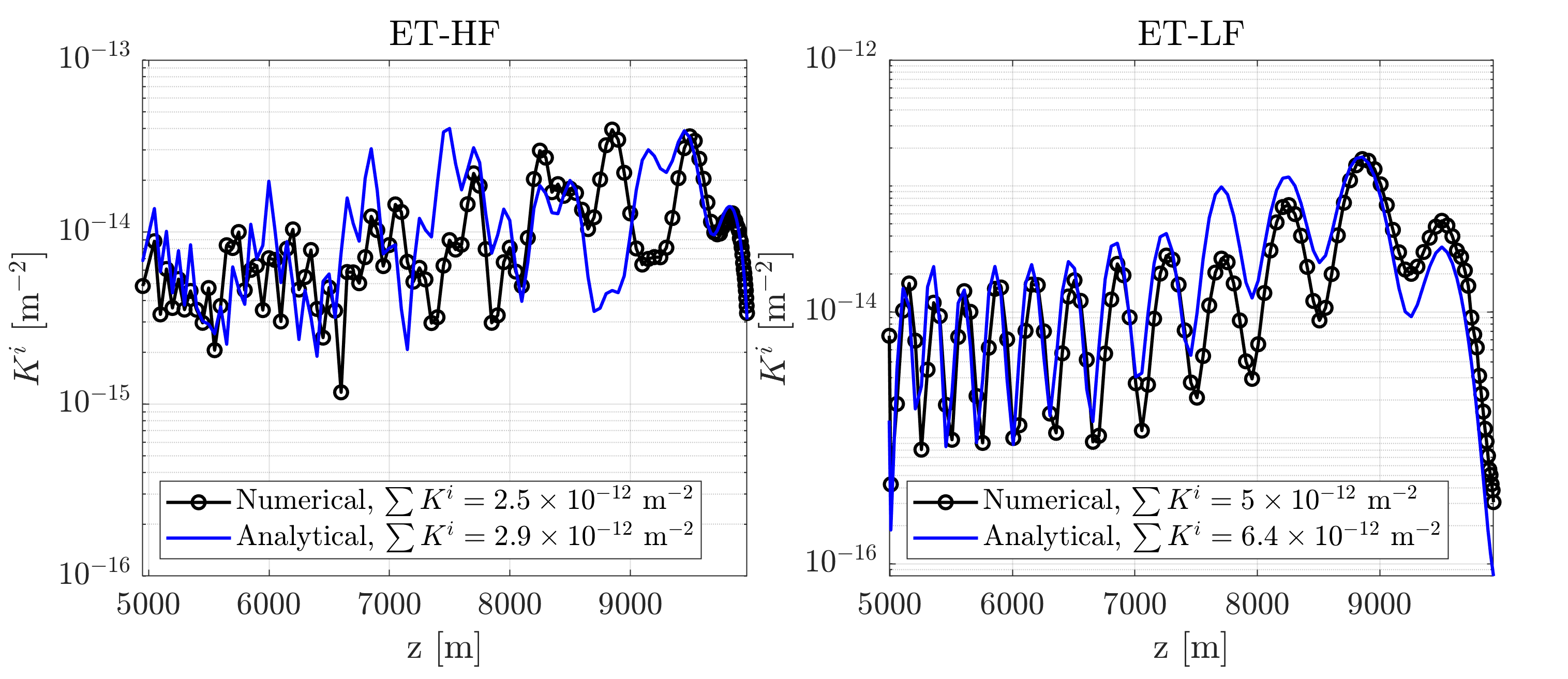}
    \caption{
        The $K^i$ values as a function of baffle position $z_i$ as computed for the different baffles at the far end of the tube using the analytical BRDF  and the numerical simulations (see body of the text) for (left) ET-HF and (right) ET-LF.}
    \label{fig:Ki}
\end{figure}

In the case of the backscattering contributions from the baffles, we use both the 
analytical BRDF and the numerical simulation tools described above to compute the power reaching the baffles. 
Following Eqs.(29) and (30), the terms $K^i$ are computed using the baffles corresponding to the far half of the tube with respect to the given mirror. The results are presented in Fig.~\ref{fig:Ki}. As anticipated, the ET-LF result displays the oscillations $\propto J_1^2(2\pi \theta R_m)/\theta^2$ sensitive to the finite size of the mirror,  whereas for ET-HF the effect is less pronounced,  indicating the presence of important contributions from mirror surface aberrations. At very large $z$, the values for $K^i$ decrease 
significantly.  The scattered light power distribution increases at low angles and the baffles far away from the mirror should be exposed to much more light. However, they are shielded by the presence of the rest of the baffles in the vacuum tube resulting into a very small contribution to the noise.  

We sum all the $K^i$ values for different $z_i$ and multiply the result by two in order to effectively account for the whole cavity.  In the case of ET-HF,  we obtain values for $\sum_{i=1}^{N_B} K^i$ equal to $2.9\times 10^{-12}$ m$^{-2}$ using the analytical approach and $2.6\times 10^{-12}$ m$^{-2}$ from numerical simulations.  Similarly, in the case of ET-LF,  we obtain $6.4\times 10^{-12}$ m$^{-2}$ using the analytical approach and $5.0\times 10^{-12}$ m$^{-2}$ from numerical simulations.  The observed 10$\%$ to $30\%$ difference between analytical and numerical results  
are expected given the assumptions adopted in the analytical estimations and the uncertainties in the modeling of the scattered light and its contribution to the noise in the interferometer. 

\begin{figure*}[htb]
    \centering
    \includegraphics[width=\columnwidth]{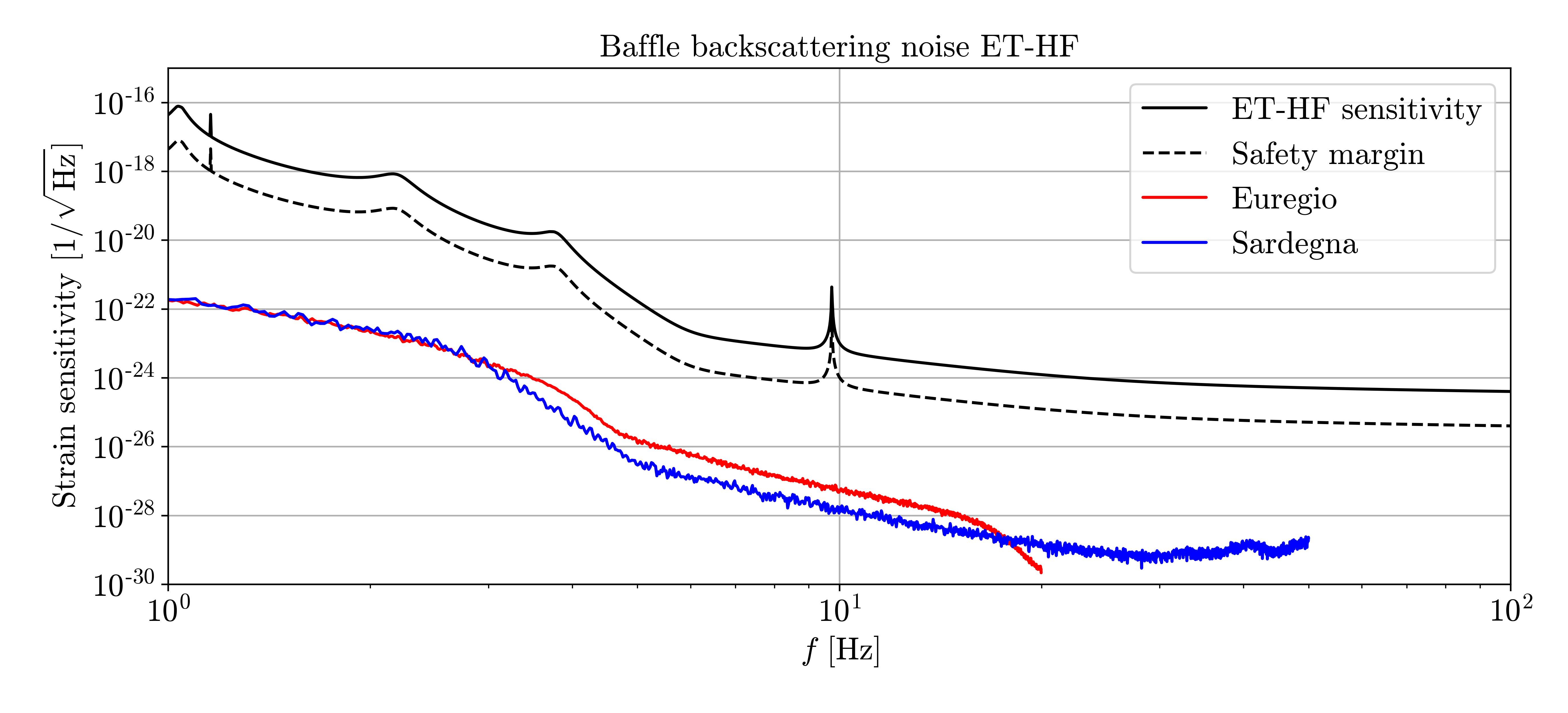}
    \includegraphics[width=\columnwidth]{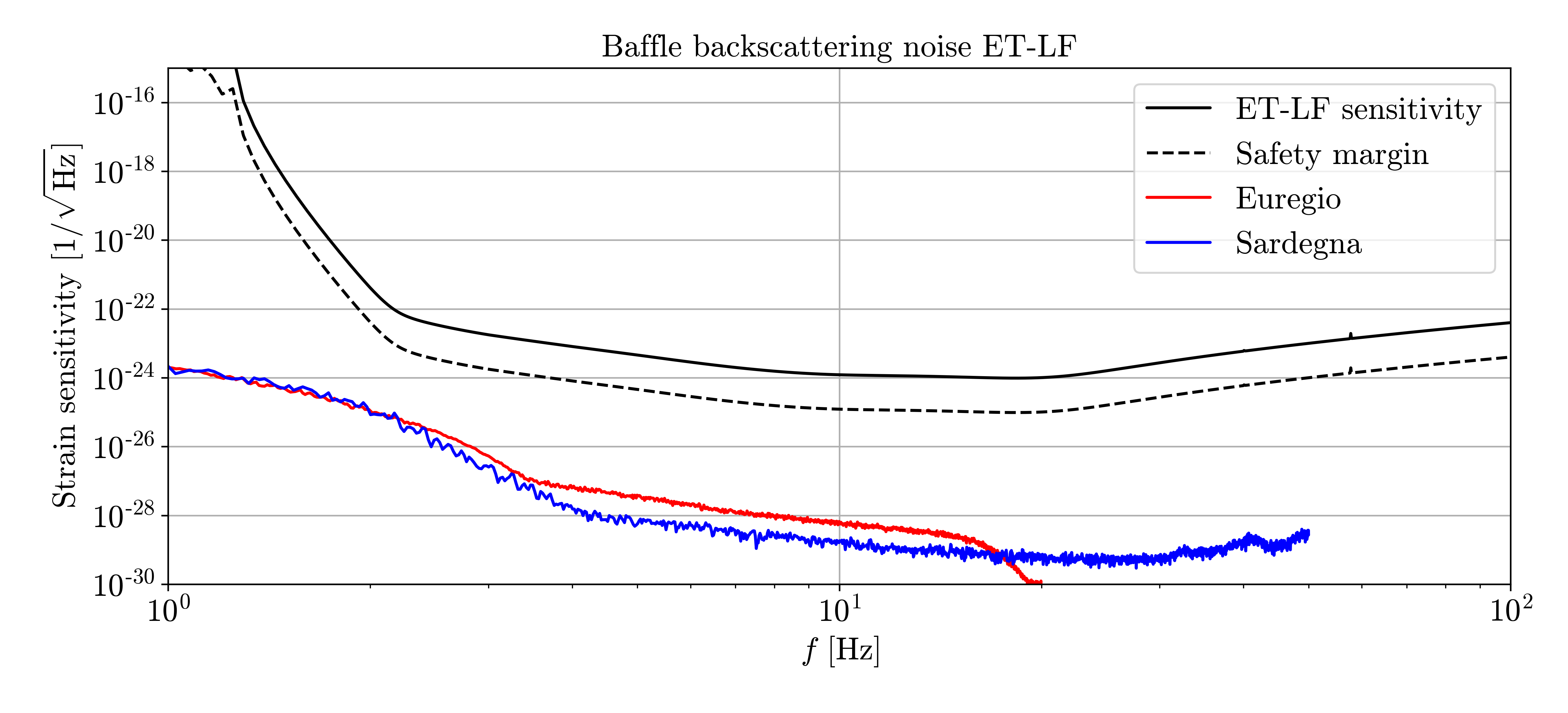}
    \caption{Stray light noise due to backscattering effects as a function of frequency. The results are computed using the seismic noise data from the Euregio (red lines) and the Sardegna (blue lines)  sites and  are compared to the anticipated ET sensitivity (black lines) and the corresponding 1/10 safety margin (dashed lines). The results are presented separately for (left) ET-HF and (right) ET-LF configurations.}
    \label{fig:NoiseETHF}
\end{figure*}

Figure~\ref{fig:NoiseETHF} shows the resulting backscattering noise separately for ET-HF and ET-LF configurations.  In both cases, the predicted noise curve is more than one order of magnitude below the expected ET sensitivity~\cite{NewETsensitivity}, indicating that the scattered light noise in the 
main cavities, dressed with baffles, should not be a limiting factor for the sensitivity of the experiment. 

The backscattering noise is recomputed in appendix~\ref{sec:App12} for a modified ET-HF configuration with a larger vacuum tube radius and larger baffle apertures, for which the induced noise is further reduced.  Finally, appendix~\ref{sec:large_mirror} considers the scenario of a possible ET-LF upgrade with larger mirrors with slightly different radius of curvature, which translates into an increase of about a factor of two in the induced backscattering noise.

\begin{figure*}[htb]
    \centering
    \includegraphics[width=\columnwidth]{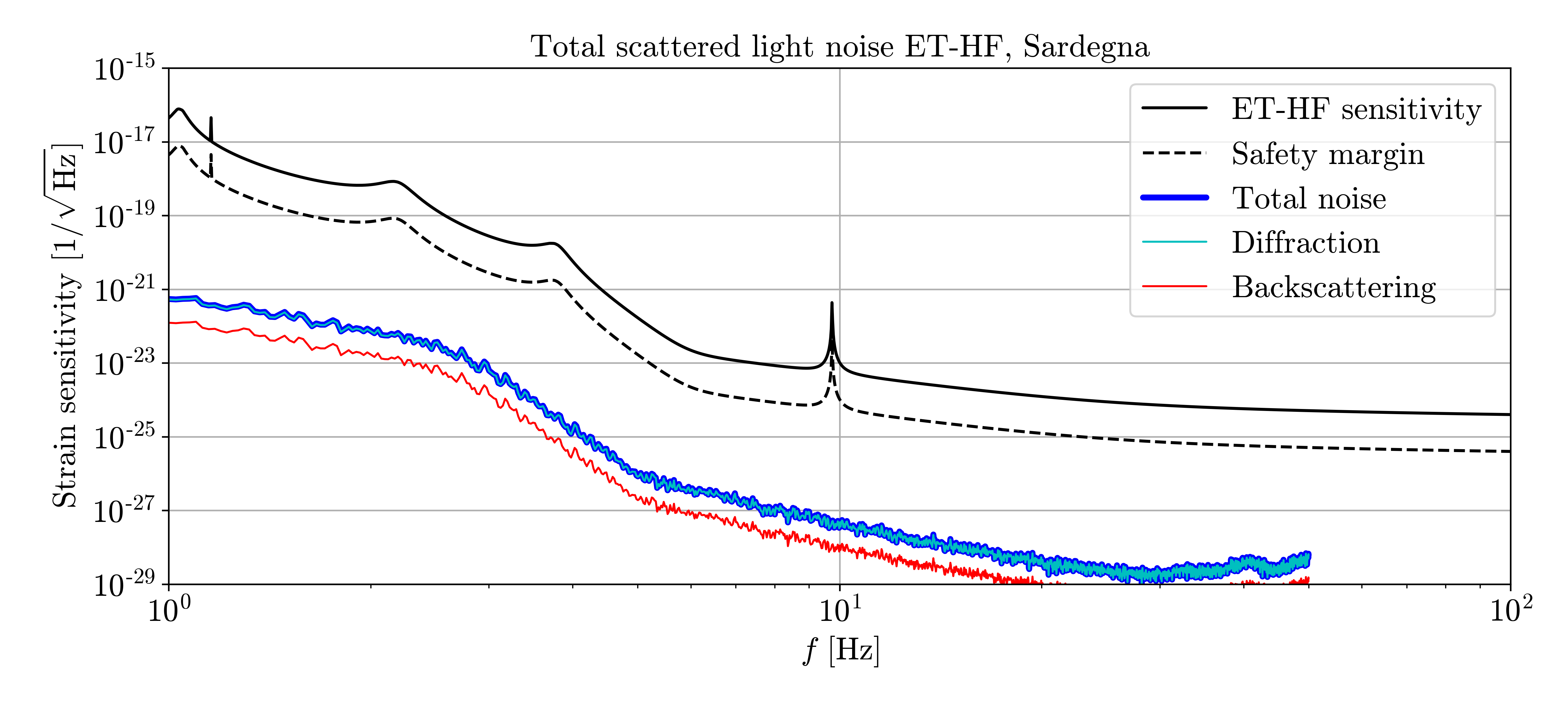}
    \includegraphics[width=\columnwidth]{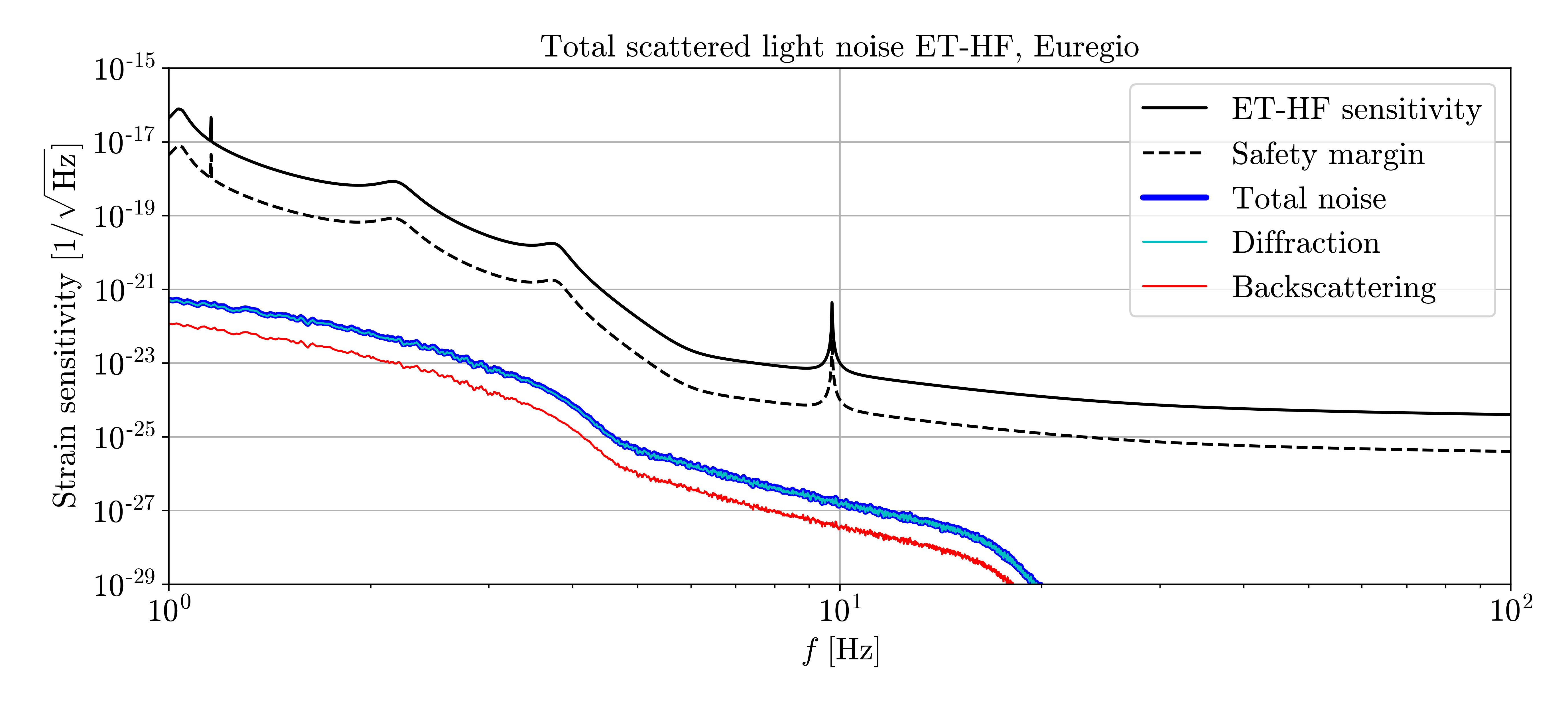}\\
    \includegraphics[width=\columnwidth]{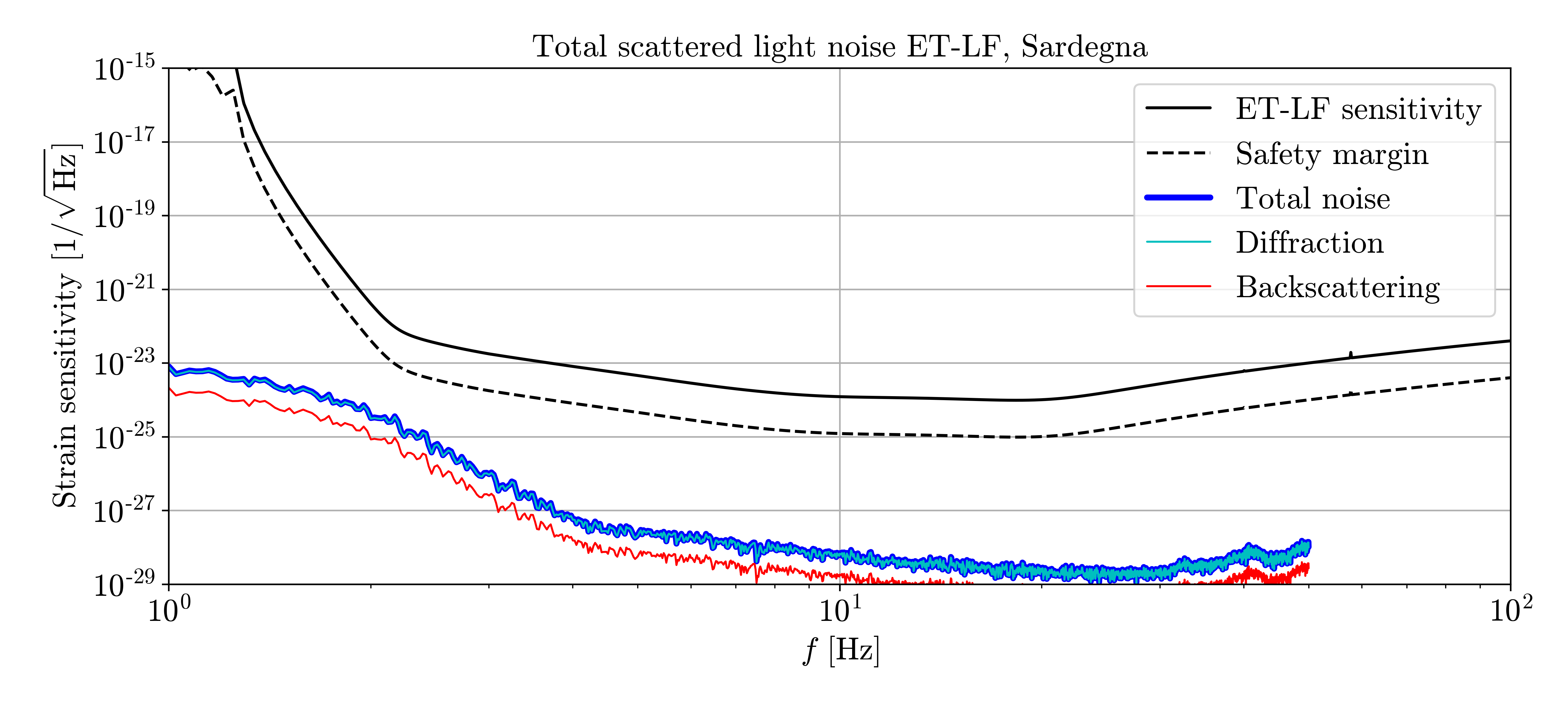}
    \includegraphics[width=\columnwidth]{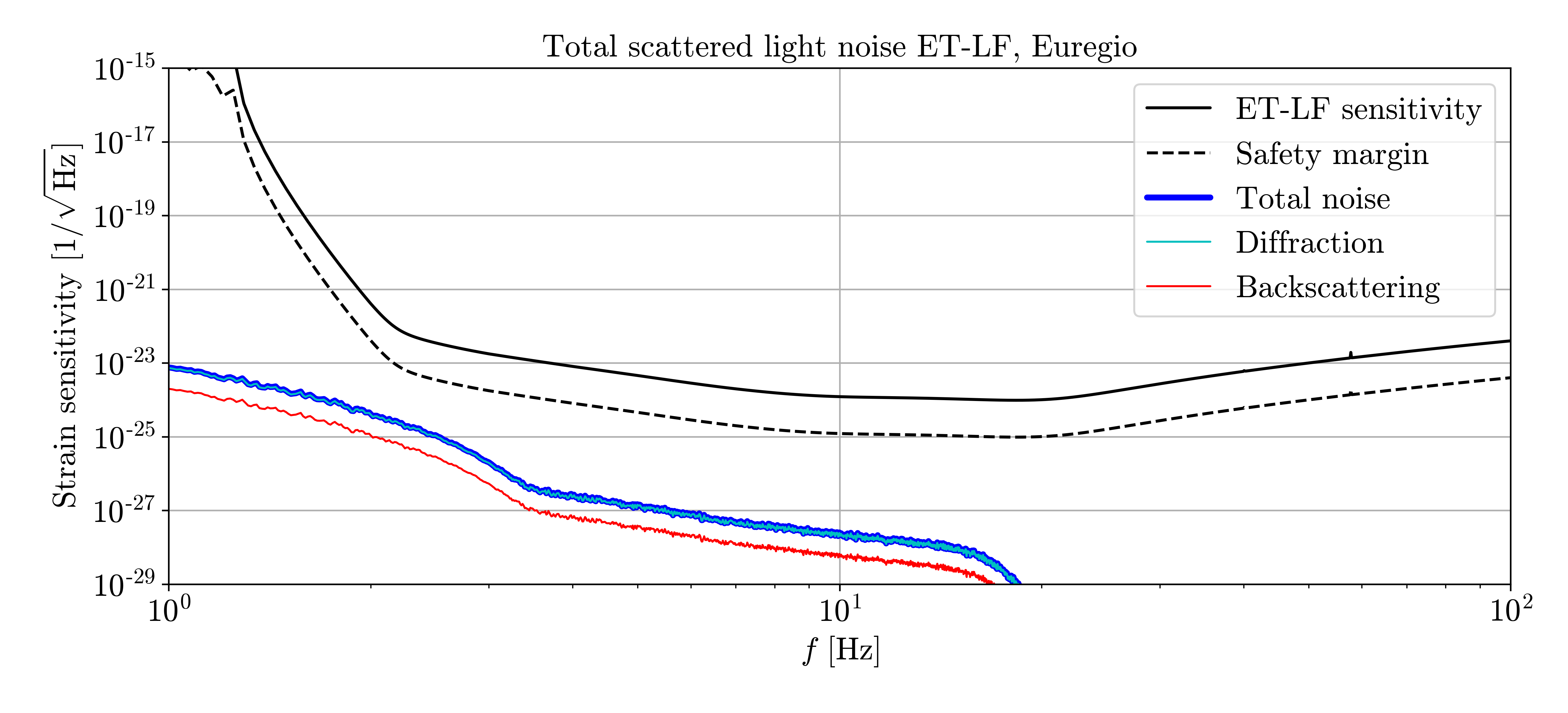}
    \caption{
    Stray light noise due to diffraction effects (cyan lines), backscattering effects (red lines), and the total noise (blue lines) as a function of frequency compared to the anticipated (top) ET-HF  and (bottom) ET-LF sensitivity curves (black lines) and the corresponding 1/10 safety margin (dashed lines). The results are computed using the seismic noise data from (left) the Sardegna site and (right) the Euregio site.}
    \label{fig:total_LF_SA}
\end{figure*}

The final results including both diffraction and backscattering related contributions are shown in Fig.~\ref{fig:total_LF_SA},  separately for ET-LH, ET-HF and for each  
site.  We conclude that, in all cases and for the baffle layout strategy proposed inside the main vacuum pipes, the expected noise levels remain well below the expected Einstein Telescope sensitivity.  

\section{Conclusions and outlook}

We present a comprehensive study of the stray light noise contributions expected in the 
vacuum pipe for the main arms of the low- and high-frequency nominal configurations of the Einstein Telescope. 
Following a similar approach already implemented in previous experiments, we discuss a noise  mitigation strategy based on the installation of conical baffles inside the vacuum pipe. For a  configuration with the laser beam centered in the cavity,  we discuss the results in terms of the 
baffle positions inside the tube, the number of baffles required  and their inner apertures.  We compute the dominant contributions to the stray light noise from diffraction by the baffle apertures and the backscattering in the baffle surfaces.  In the case of the diffraction-related noise, it is is reduced to 
a negligible level once randomly serrated conical baffles are implemented. In the case of backscattering noise contributions, we performed a detailed study using analytical expressions  and numerical simulations.  The calculations depend on a number of parameters related to the optical quality of the mirrors, the expected seismic noise levels, and the optical quality of the baffles implemented for which we use default values that can be considered conservative.  We also explore the effect of an increased beam pipe dimension and the implementation of larger mirrors in the high- and low-frequency configurations, respectively.  We conclude that, for the baffle layout strategy proposed inside the main vacuum pipes, the expected backscattering noise levels remain well below the expected Einstein Telescope sensitivity.  

Our studies leave room for the optimization of the different elements dictating the scattered light noise. As an example, larger than expected seismic noise levels or particularly large light exposures can be locally mitigated by the implementation of dampers or suspended baffles, and the noise budget can be easily reduced by a factor of two by increasing the quality of the polishing and/or anti-reflective coating of the baffle surfaces. Moreover, the interface with other details 
related to the vacuum system will condition also the exact location of baffles inside the pipes. 

Work is in progress to determine the scattered light noise levels in other scenarios including: significant laser beam offsets, mis-alignments, and the presence of defects and point absorvers in the mirrors.  Finally, the scope of this  study does not include 
crucial considerations on the stray light background levels originated 
in the vicinity of the main mirrors, inside the cryotrap areas attached to the vacuum towers hosting the mirror's suspensions, or inside the vacuum towers themselves. This is the subject of a separate study.

\section*{Acknowledgements}

We thank A. Ananyeva, L. Conti, A. Grado, H. Lueck  and other member of the ET Collaboration for their contributions to the discussion of the results. 
This
project has received funding from the European Union’s Horizon 2020 research and innovation programme under the Marie Skłodowska-Curie Grant Agreement No. 754510.
This work
is partially supported by the Spanish MCIN/AEI/10.13039/501100011033 under the Grants
No. SEV-2016-0588, No. PGC2018-101858-B-I00, and No. PID2020-113701GB-I00, some
of which include ERDF funds from the European Union, and by the MICINN with funding from the European Union NextGenerationEU (PRTR-C17.I1) and by the Generalitat de
Catalunya. IFAE is partially funded by the CERCA program of the Generalitat de Catalunya.
MAC is supported by the 2022 FI-00335 grant. Part of this material is based upon work supported by NSF’s LIGO Laboratory which is a major facility fully funded by the National Science Foundation and operates under Cooperative Agreement No. PHY-1764464.

\appendix
\section{Effect of increasing the tube radius of ET-HF}\label{sec:App12}
We study the effect in terms of the scattered light noise levels of enlarging the tube radius for the ET-HF interferometer. This is motivated by the fact that a larger aperture might provide an additional margin for
operating the interferometer with large laser offsets in the arms. It will also mitigate the risk of 
larger than expected backscattering noise due to the eventual presence of unforeseen(and difficult to model) 
point absorbers, point scatterers, or mirror defects that might change significantly the mirror BRDF and 
increase the amount of light illuminating the baffles. We consider an enlarged vacuum tube inner radius 
$R = 0.6$~m and a baffle aperture of $A_b = 1.04$~m.  As shown in Fig.~\ref{fig:NumOfBafflesBafflesHF}, this new configuration does not translate into a very significant increase in the number of baffles installed in the arms. 

\begin{figure}[htb]
    \centering
    \includegraphics[width=\columnwidth]{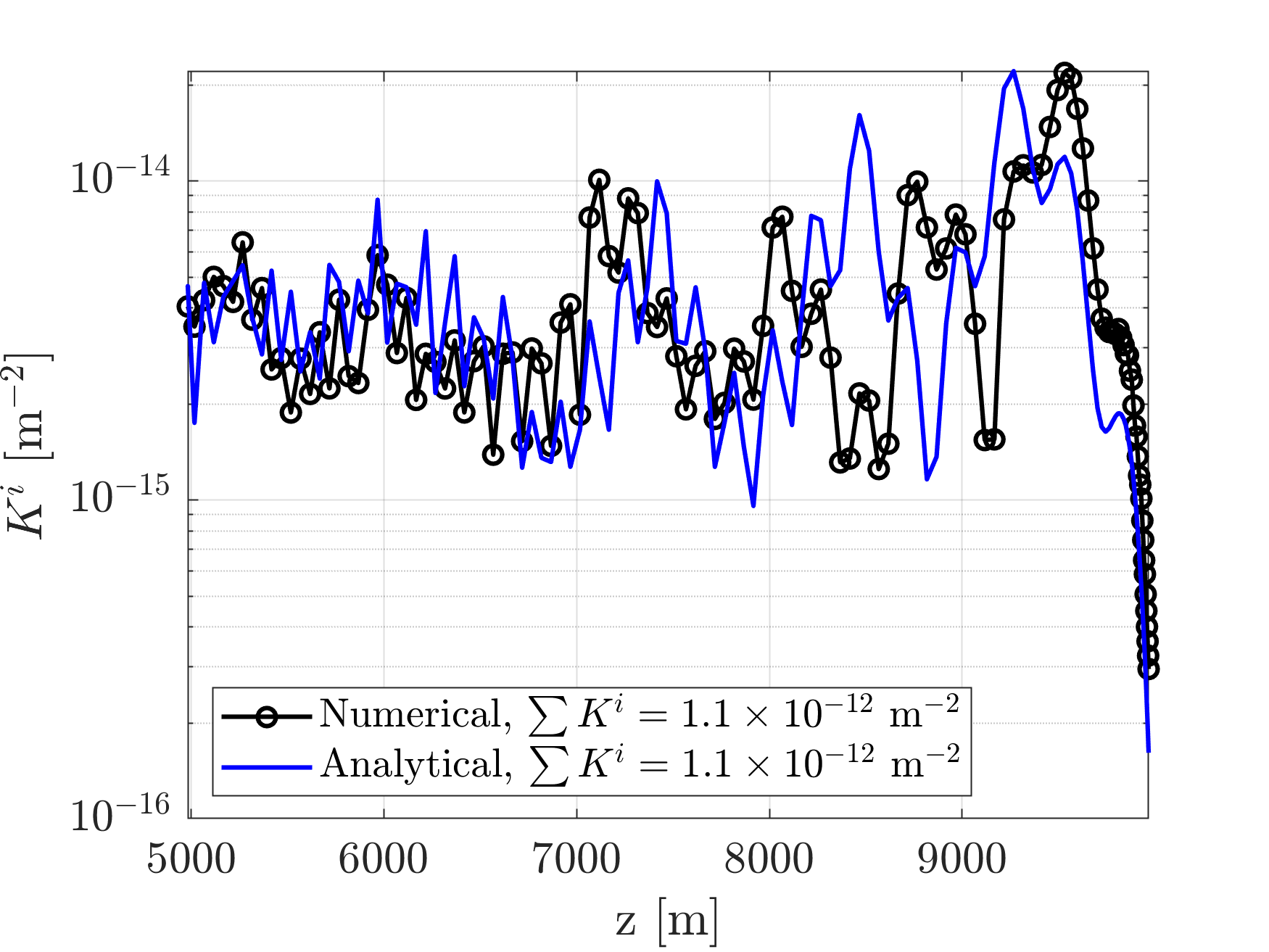}
    \caption{
    The $K^i$ values as a function of baffle position $z_i$ as computed for the different baffles at the far end of the tube using the analytical model of the BRDF and the numerical simulations (see body of the text) for ET-HF and a vacuum tube radius of 0.6~m.}
    \label{fig:Ki_12}
\end{figure}

\begin{figure}[htb]
    \centering
    \includegraphics[width=\columnwidth]{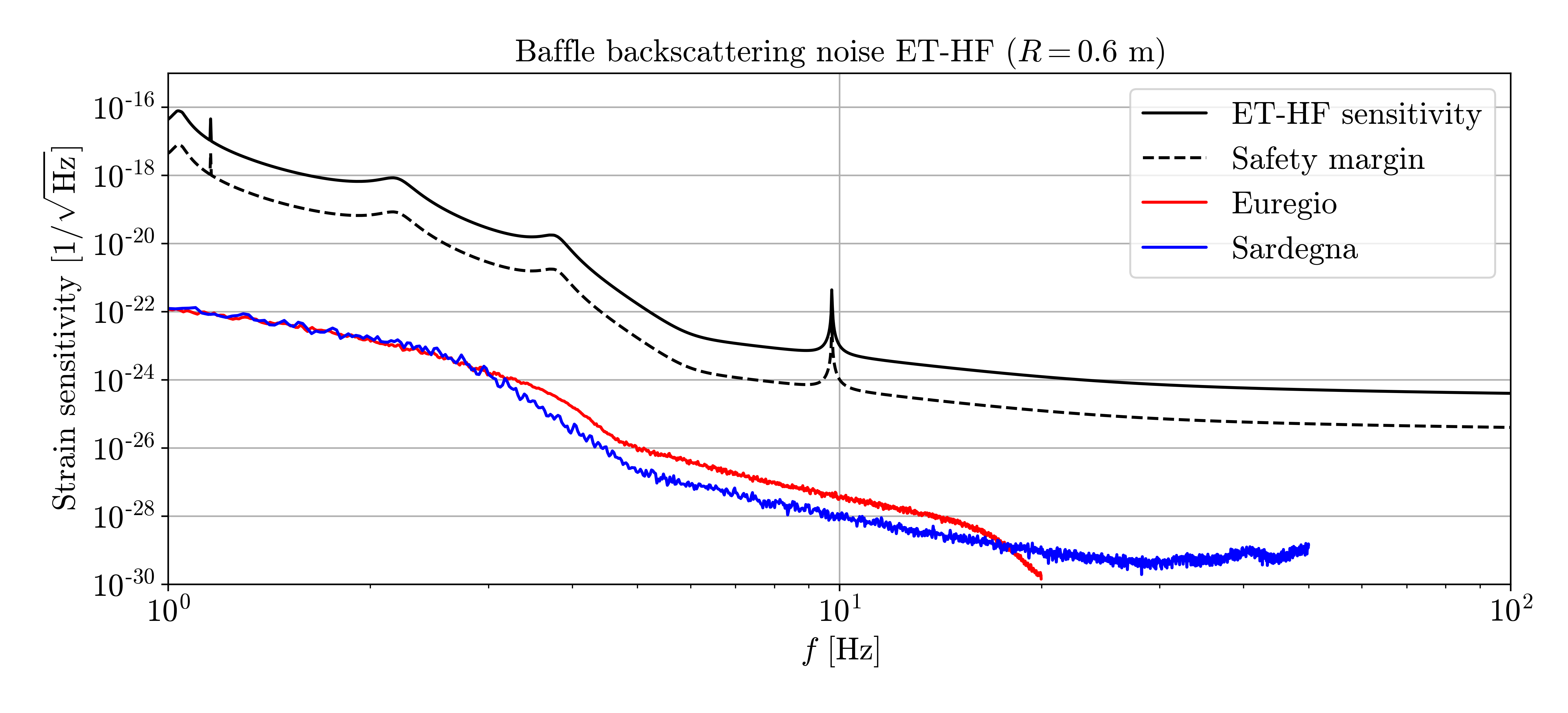}
    \caption{
    Stray light noise due to backscattering effects as a function of frequency for 0.6~m vacuum tube radius. The results are computed using the seismic noise data from the Euregio (red line) and the Sardegna (blue line)  sites and  are compared to the anticipated ET-HF sensitivity (black line) and the corresponding 1/10 safety margin (dashed line).}
        \label{fig:NoiseETLF_12}
\end{figure}

In terms of the calculation of the backscattering noise from the baffles, the new analytical  and simulated estimations of $K^i$ are displayed in Fig. \ref{fig:Ki_12}. After including all the contributions from each individual baffle we obtain $\sum K^i= 1.1\times 10^{-12}$ m$^{-2}$. As expected, the larger aperture results in a large reduction in the scattered light arriving to the baffles.  As presented in Fig.~\ref{fig:NoiseETLF_12},  it turns into a further reduction of about  
a factor two in the predicted noise.  

\section{Effect of the finite mirror dimension in the ET-LF BRDF}\label{sec:finite}

In order to illustrate the effect of the a finite mirror dimension in the  mirror BRDF, we consider a perfect surface (i.e. no aberration) but finite in size, which will have the following mirror map

\begin{equation} \label{eq:MirrorMap}
    \delta(x,y) = \left\{
    \begin{array}{lr}
        \displaystyle h, & \text{if }x^2+y^2\leq R_m^2\\ \\
        \displaystyle 0, & \text{if } x^2+y^2> R_m^2
    \end{array}
\right.~.
\end{equation}

The two-dimensional Fourier transform, defined as
\begin{equation}
    F(f_x,f_y) = \int_{-\infty}^{+\infty}\int_{-\infty}^{+\infty} \delta(x,y)e^{i2\pi (xf_x+yf_y)}\td x \td y~,
\end{equation}
\noindent
can be simplified if there is circular symmetry and is expressed in polar coordinates, 
\begin{equation}
    F(\xi,\varphi) = 2\pi \int_{0}^\infty r\delta(r)J_{0}(2\pi\xi r)\td r~,
\end{equation}
where $J_0$ is the zeroth Bessel function of first kind. Evaluating this integral for the mirror map defined in Eq. \ref{eq:MirrorMap} yields

\begin{equation}
    F(\xi,\varphi) = h R_m \frac{J_1(2\pi \xi R_m)}{\xi}~,
\end{equation}
where the property 
\begin{equation}
    \int_0^\eta x J_0(x)\td x = \eta J_1(\eta)
\end{equation}
has been used. Since the PSD is defined as $S(\xi,\varphi) = |F(\xi,\varphi)|^2$, it can be stated that
\begin{equation}
    S(\xi,\varphi) \propto \frac{J_1^2(2\pi\xi R_m)}{\xi^2}  
\end{equation}
\noindent 
and 
\begin{equation}
\mathrm{BRDF}\propto \frac{J_1^2(2\pi\theta R_m / \lambda)}{\theta^2}~, 
\end{equation}
which is the pattern observed in Fig. \ref{fig:BRDF} for the ET-LF interferometer.

%
%

\section{Effect of larger mirrors in ET-LF}
\label{sec:large_mirror}
We compute the stray light noise in an ET-LF configuration with larger mirrors than anticipated in the 
current design.  This is motivated by future potential ET upgrades targeting the reduction 
of the thermal noise in the mirrors. Here we consider new mirrors with a radius of $31$~cm and modified radius of curvature.   
Using the criteria of maintaining the same level of 
clipping losses due to the finite size of the mirror to about $3.63$ ppm,  the new radius of curvature is computed. If both mirrors have the same radius of curvature, this is related to the rest of the parameters of the cavity~\cite{Kogelnik66}
\begin{equation}
    \mathcal{R} = \frac{2\pi^2 w_0^4}{\lambda^2 L}+\frac{L}{2}~.
\end{equation}
\noindent
To determine the beam waist, we use Eq.~\ref{eq:w(z)} to relate
it to the beamspot in the mirror

\begin{equation}
    w(0)^2=2w_0^2+\frac{L^2\lambda^2}{4\pi^2w_0^2}~,
\end{equation}
\noindent
with solutions
\begin{equation}
    w_0^2=\frac{1}{2}\left( w(0)^2\pm \sqrt{w(0)^4-\frac{L^2\lambda^2}{\pi^2}}\right)~,
\end{equation}
\noindent
where the negative one gives the correct beam waist.

The beamspot at a given level of clipping losses can be computed using Eq. \ref{eq:r_Lc}.
Finally, the radius of curvature is determined as 
\begin{equation}
    \mathcal{R}= \frac{\pi^2}{2\lambda^2L}\left[ \frac{2R_m^2}{\ln(1/L_c)} - \sqrt{\frac{4R_m^4}{\ln^2(1/L_c)}-\frac{L^2\lambda^2}{\pi^2}}\right]^2+\frac{L}{2}~.
\end{equation}
\noindent
Evaluating this expression for the new radius of the mirrors and the same clipping losses leads to a radius of curvature of $\mathcal{R}=5136$ m.  

\begin{figure}[htbp]
    \centering
    \includegraphics[width=\columnwidth]{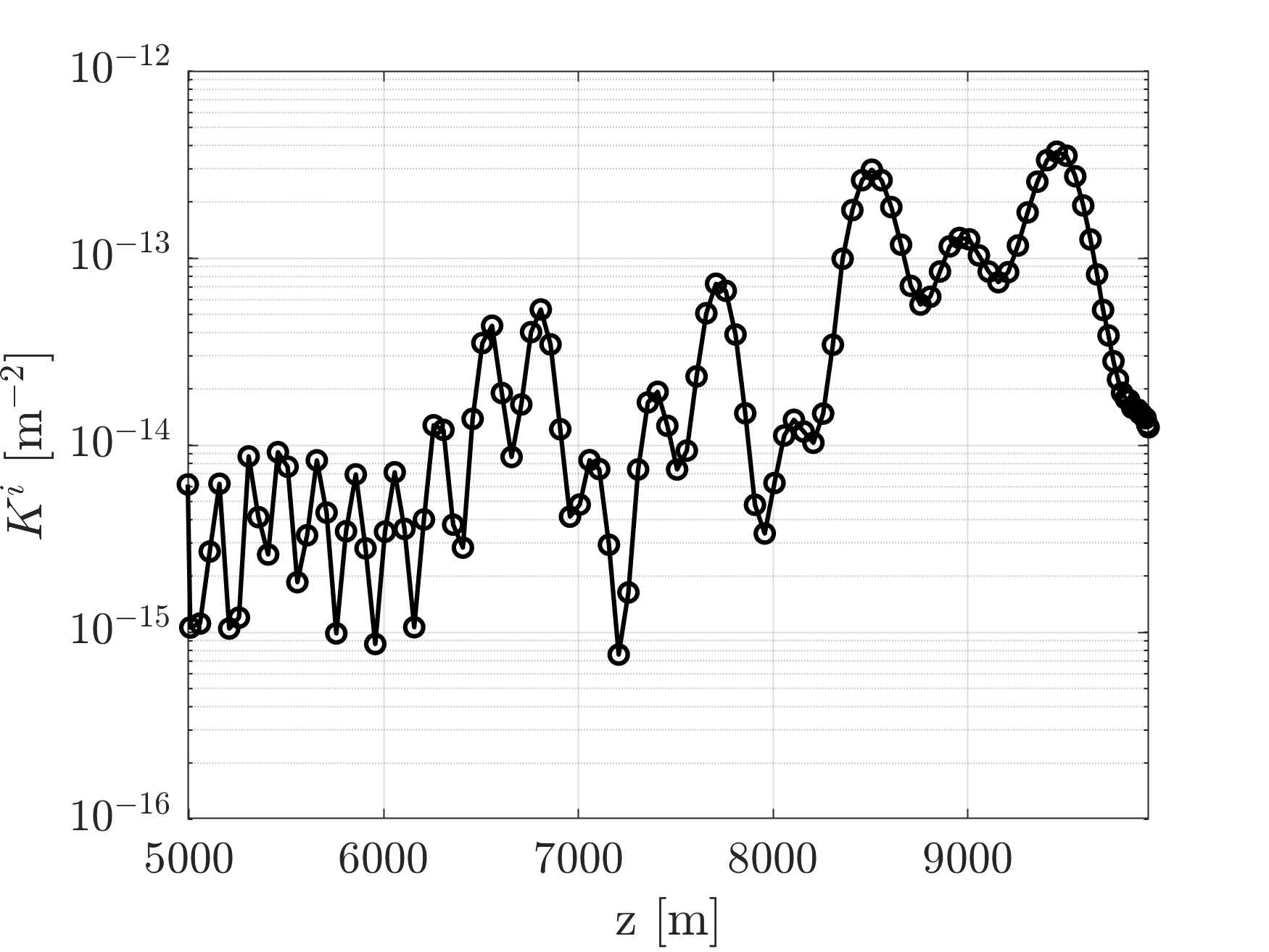}
    \caption{
    The $K^i$ values as a function of baffle position $z_i$ as computed for the different baffles at the far end of the tube using numerical simulations (see body of the text) for ET-LF and considering larger mirrors.}
        \label{fig:Ki_ETLF}
\end{figure}

\begin{figure}[htbp]
    \centering
    \includegraphics[width=\columnwidth]{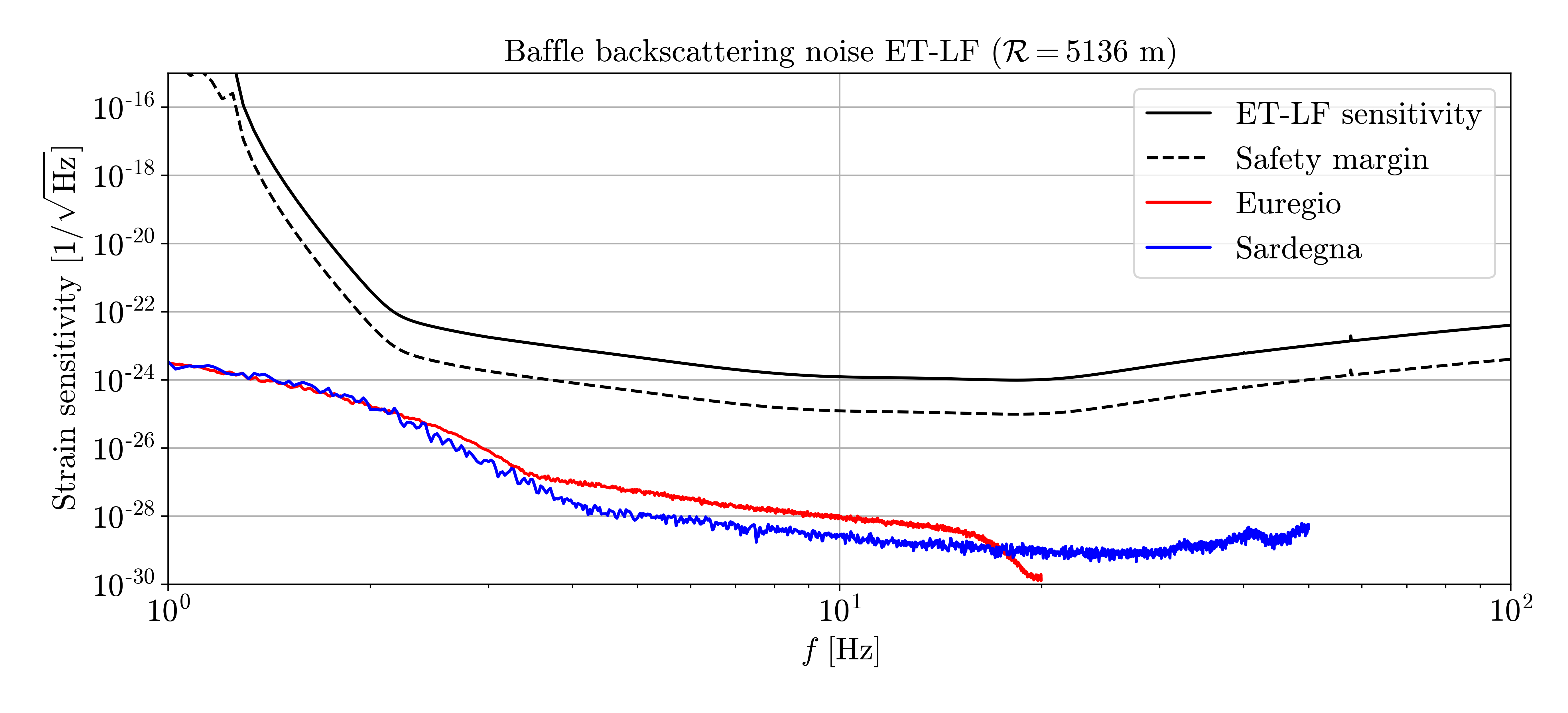}
    \caption{
Stray light noise due to backscattering effects as a function of frequency for a larger mirror size with 31~cm radius. The results are computed using the seismic noise data from the Euregio (red line) and the Sardegna (blue line)  sites and  are compared to the anticipated ET-LF sensitivity (black line) and the corresponding 1/10 safety margin (dashed line).}
    \label{fig:Noise_ET_LFroc}
\end{figure}

The new results for the $K^i$ distributions, using the new cavity, are presented in Fig. \ref{fig:Ki_ETLF} and correspond to a total of $\sum K ^i = 1.2  \times 10^{-11}$ m$^{-2}$.
This represents about a factor two increase in the backscattering noise contribution, which 
still remains well below the anticipated ET-LF sensitivity,  as shown in Fig. \ref{fig:Noise_ET_LFroc}. 

\bibliographystyle{unsrtnat}
\bibliography{ref.bib}{}


\end{document}